\documentclass[aps,twocolumn]{revtex4-1}
\usepackage{times}
\usepackage{graphicx}
\usepackage{amsfonts}
\usepackage{amsmath, amsthm, amssymb}
\usepackage{dsfont}
\usepackage{color}
\usepackage{empheq}
\usepackage{standalone}
\usepackage[most]{tcolorbox}
\usepackage{physics}
\usepackage{bm}
\usepackage[utf8]{inputenc}
\usepackage{amssymb,amsmath,amsfonts,amscd}
\usepackage{mathrsfs}
\usepackage{siunitx}
\usepackage{commath}
\usepackage{graphicx}       
\graphicspath{ {Images/} }  
\usepackage{braket}
\usepackage{bm}
\usepackage{empheq}
\usepackage[most]{tcolorbox}
\usepackage{physics}
\usepackage{hyperref} 
\hypersetup{
    colorlinks=true,
    linkcolor=black,    
    citecolor=blue,    
    urlcolor=blue,      
}

\begin{document}

\title{Magnon-mediated superconductivity on the surface of a topological insulator}

\author{Eirik Erlandsen}
\affiliation{\mbox{Center for Quantum Spintronics, Department of Physics, Norwegian University of Science and Technology,}\\NO-7491 Trondheim, Norway}
 
 \author{Arne Brataas}
\affiliation{\mbox{Center for Quantum Spintronics, Department of Physics, Norwegian University of Science and Technology,}\\NO-7491 Trondheim, Norway}

\author{Asle Sudbø}
\email[Corresponding author: ]{asle.sudbo@ntnu.no}
\affiliation{\mbox{Center for Quantum Spintronics, Department of Physics, Norwegian University of Science and Technology,}\\NO-7491 Trondheim, Norway}
 

\begin{abstract}
We study superconductivity on the surface of a topological insulator, mediated by magnetic fluctuations in an adjacent ferromagnetic or antiferromagnetic insulator. Superconductivity can arise from effective interactions between helical fermions induced by interfacial fermion-magnon interactions. For both ferromagnetic and antiferromagnetic insulators, these fermion-fermion interactions have the correct structure to facilitate pairing between particles located on the same side of the Fermi surface, also known as Amperean pairing. In antiferromagnets, the strength of the induced interactions can be enhanced by coupling the topological insulator asymmetrically to the two sublattices of the antiferromagnet. This effect is further amplified by next nearest neighbor frustration in the antiferromagnetic insulator. The enhancement makes the induced interactions significantly stronger in the antiferromagnetic case, as compared to the ferromagnetic case. These results indicate that an uncompensated antiferromagnetic interface might be a better candidate than a ferromagnetic interface for proximity-induced magnon-mediated superconductivity on the surface of a topological insulator.
\end{abstract}

\pacs{Valid PACS appear here}

\maketitle

\section{Introduction}
{Heterostructures of ferro- and antiferromagnetic insulators on the one hand, and superconductors, metals, and topological insulators on the other hand, have received much interest both theoretically and experimentally over the last decades \cite{Tserkovnyak2002, Buzdin2005, Saitoh2006, Linder2010, Garate2010, Wei2013, Cheng2014, Cornelissen2015, Du2017, Lebrun2018}. They continue to be fruitful model systems for development of novel ideas in condensed matter physics. Recently, the idea that magnons in a magnetic material can induce superconductivity across an interface when proximized with either a normal metal (NM) or a topological insulator (TI) has been considered in some detail \cite{Kargarian2016, Gong2017, Fjaerbu2018, Hugdal2018, Fjaerbu2019, Erlandsen2019}. In NMs, the magnons mediate attractive interactions between electrons. On the surface of TIs the Cooper pairs are formed by helical fermions where the spin and momentum are locked together \cite{Hasan2010, Qi2011}. As a consequence, while the pairing in NMs is of the normal BCS-type where Cooper pairs are formed by electrons on opposite sides of the Fermi surface, Kargarian \textit{et al}. \cite{Kargarian2016} predicted pairing between fermions with momenta in the same direction, named Amperean pairing \cite{Lee2007}, in a TI coupled to a ferromagnetic insulator (FMI). \color{black} Amperean pairing, with finite-momentum Cooper pairs, should be experimentally distinguishable from normal BCS-type pairing through its nonuniform ground state \cite{Kargarian2016}. \color{black} In a subsequent related study, the cases of a TI coupled to a FMI and a TI coupled to an antiferromagnetic insulator (AFMI) were considered in Ref. \cite{Hugdal2018}. Possible attractive interactions for both Amperean and BCS-type pairing were found.\\\indent}
Similarly to Kargarian \textit{et al}.\!, we study a TI coupled to a FMI. Instead of a continuum action model, we utilize a lattice model Hamiltonian to describe the system. \color{black} Furthermore, Kargarian \textit{et al}.\! applied a self-consistent strong-coupling approach, establishing that the fermionic states on the surface of the TI can be strongly renormalized by the presence of the magnetic fluctuations through the fermion self-energy. Our objective is, however, not to perform an optimal analysis of the superconducting instability. Instead, we seek to reveal the qualitative difference between the effective fermionic interactions induced by ferromagnetic and antiferromagnetic fluctuations, arising from the magnon coherence factors not present in the ferromagnet. To illustrate this important aspect, we therefore find it sufficient to apply a simpler weak-coupling approach, neglecting the renormalization of the fermionic normal state. \color{black} \\\indent 
Treating the magnetic subsystem in a quantum mechanical fashion, we investigate the effect of coupling the TI symmetrically or asymmetrically to the two sublattices of the AFMI. An asymmetric coupling can be achieved through a fully uncompensated antiferromagnetic interface where only one of the two sublattices is exposed \cite{Kamra2017B, Kamra2018}. Such an asymmetric coupling has been predicted to significantly enhance the critical temperature for magnon mediated superconductivity in a NM/AFMI heterostructure \cite{Erlandsen2019}. This enhancement can be understood from the picture of antiferromagnetic magnons as squeezed states, revealing that an antiferromagnetic magnon is associated with a large spin located at each sublattice \cite{Kamra2019}. Coupling to only one of the two sublattices of an AFMI thereby involves coupling to a large spin, leading to a strong enhancement of the coupling interaction \cite{Erlandsen2019, Johansen2019}.\\\indent 
For both FMIs and AFMIs, we find that the effective fermion-fermion interactions mediated by magnetic fluctuations cannot facilitate BCS-type chiral p-wave pairing. For Amperean pairing, on the other hand, we find that the effective potential has the correct form to produce a non-trivial solution to the gap equation in both the FMI and AFMI cases. For the FMI, the phase space is, within our weak-coupling approach, too small to produce a superconducting instability for realistic parameters, in contrast to the strong-coupling result of Kargarian \textit{et al}. For the AFMI, when coupling asymmetrically  to the two sublattices, we obtain a non-trivial solution to the gap equation. However, this solution arises from a strong interaction potential and small phase space. \color{black} Combined with the strong renormalization of the fermionic states predicted by Kargarian \textit{et al}., this suggests that a strong-coupling approach, not performed here, is needed in order to provide stronger evidence for the existence of a superconducting instability and realistic estimates for the critical temperature. \color{black}\\\indent
The main result of the paper, which is expected to be robust, is that the strength of the fermion-fermion interaction mediated by antiferromagnetic magnons, analogously to the NM/AFMI case of Ref.\! \cite{Erlandsen2019}, is enhanced by coupling asymmetrically to the two sublattices of the AFMI. The interaction strength is therefore significantly larger than for an FMI. Moreover, including an antiferromagnetic next nearest neighbor interaction term, frustrating the AFMI, is found to strengthen the enhancement effect. The result of this next nearest neighbor frustration is not limited to the case of a TI.\\
\indent The paper is organized as follows. In Sec. \ref{Section:FMI_TI} we consider the case of a TI coupled to a FMI, and in Sec. \ref{Section:AFMI_TI} we consider the case of a TI coupled to an AFMI. The results are summarized in Sec. \ref{Section:Summary}. Additional details concerning the derivation of the self-consistent equation for the Amperean gap function are presented in the Appendix. 

\section{Ferromagnetic case}\label{Section:FMI_TI}
\begin{figure}[ht] 
    \begin{center}
        \includegraphics[width=0.65\columnwidth,trim= 0.1cm 0.1cm 0.0cm 0.2cm,clip=true]{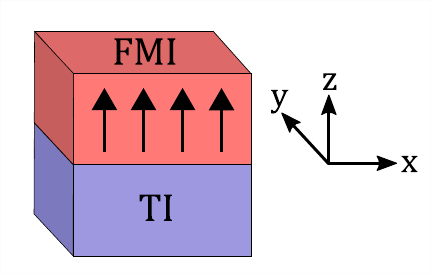}
    \end{center}
    \caption{The system consists of a ferromagnetic insulator (FMI) on top of a topological insulator (TI).}
    \label{fig:FMI_TI}
\end{figure}

We consider a 3D TI with a single Dirac cone such as $\rm{Bi}_2\rm{Se}_3$ or $\rm{Bi}_2\rm{Te}_3$ \cite{Xia2009, Chen2009} proximity-coupled to a FMI such as $\rm{YIG}$, $\rm{EuO}$ or $\rm{EuS}$ \cite{Wei2013, Li2015, Yang2013}, as displayed in Fig.\! \ref{fig:FMI_TI}. The interface is placed in the $xy$-plane. For the FMI we assume an ordered magnetic state with a magnetization along the z-axis. In the following, we take $\hbar = a = 1$, where $a$ is the lattice constant.
\subsection{Model} The system is modeled by the Hamiltonian \cite{Zhou2017, Li2014}\, $H = H_{\text{FMI}} + H_{\text{TI}} + H_{\text{int}}$,
\begin{subequations}
\begin{align}
    &H_{\rm{FMI}} = -J \!\sum_{\langle \bm{i}, \bm{j} \rangle} \bm{S}_{\bm{i}} \cdot \bm{S}_{\bm{j}} - K\sum_{\bm{i}}S^2_{\bm{i}z},\label{H_FM_F}\\
   &H_{\rm{TI}} = \frac{v_{F}}{2}\sum_{\bm{i},\bm{b}}[c_{\bm{i}}^{\dagger}(i\tau_{y}\delta_{\bm{b}, \Hat{x}} - i\tau_{x}\delta_{\bm{b}, \Hat{y}})c_{\bm{i}+\bm{b}} + \rm{h.c.}] \nonumber\\
   &+ \sum_{\bm{i}}c^{\dagger}_{\bm{i}}[2W\tau_{z} - \mu]c_{\bm{i}} - \frac{W}{2}\sum_{\bm{i},\bm{b}}[c_{\bm{i}}^{\dagger}\tau_z c_{\bm{i}+\bm{b}} + \rm{h.c.}],\label{H_TI_F}\\
    &H_{\rm{int}} =  -2\Bar{J} \sum_{\bm{i}} c_{\bm{i}}^{\dagger}\bm{\tau}c_{\bm{i}}\cdot \bm{S_i}.
    \label{H_int_F}
\end{align}
\end{subequations}
Here, the ferromagnetic exchange interaction between lattice site spins $S_{\bm{i}}$ is parametrized by the exchange constant $J > 0$, and the strength of the easy-axis anisotropy is determined by $K > 0$. Moreover, $c_{i}^{\dagger} = (c_{i\uparrow}^{\dagger}, c_{i\downarrow}^{\dagger})$, where $c_{i\sigma}^{\dagger}$ is a creation operator for an electron with spin $\sigma$ on lattice site $\bm{i}$ on the surface of the TI. The Pauli matrices $\bm{\tau}$ act on the spin degree of freedom of the electrons and $\mu$ is the chemical potential. The first term in the TI-Hamiltonian of Eq.\! \eqref{H_TI_F} describes the spin-momentum locking experienced by electrons on the surface of the TI, the strength of which is determined by the Fermi velocity $v_{F}$. The following Wilson terms (W) ensures that there is not more than one Dirac cone in the first Brillouin zone, avoiding the fermion doubling problem which arises in the discretization of the continuum model $H(\bm{k}) = v_{\rm{F}}(\bm{\tau} \cross \bm{k})\cdot \Hat{z}\,$ \cite{Zhou2017, Li2014, Kogut1983}. \color{black} The Wilson terms are phenomenologically added to the Hamiltonian in order to produce a lattice model that describes the correct physics. Their effect vanish in the long-wavelength limit where the effective 2D lattice model is expected to faithfully describe the surface states of the TI \cite{Zhou2017}. \color{black} The electrons on the surface of the TI are exchange coupled to the lattice site spins on the surface of the FMI \cite{Fjaerbu2018}, with a strength determined by $\Bar{J}$. The lattices are quadratic and we assume periodic boundary condition in the x,y directions in order to capture the physics at the interface between the two materials. The sum over ${\langle \bm{i}, \bm{j} \rangle}$ includes all nearest neighbors in both positive and negative directions, while $\bm{b}\in\{\Hat{x},\Hat{y}\}$ only includes nearest neighbors in the positive directions.
\subsection{Diagonalization of subsystems} 
\begin{figure}[ht] 
    \begin{center}
    \hspace*{-1.2cm} 
        \includegraphics[width=0.70\columnwidth,trim= 0.4cm 0.2cm 1.0cm 1.0cm,clip=true]{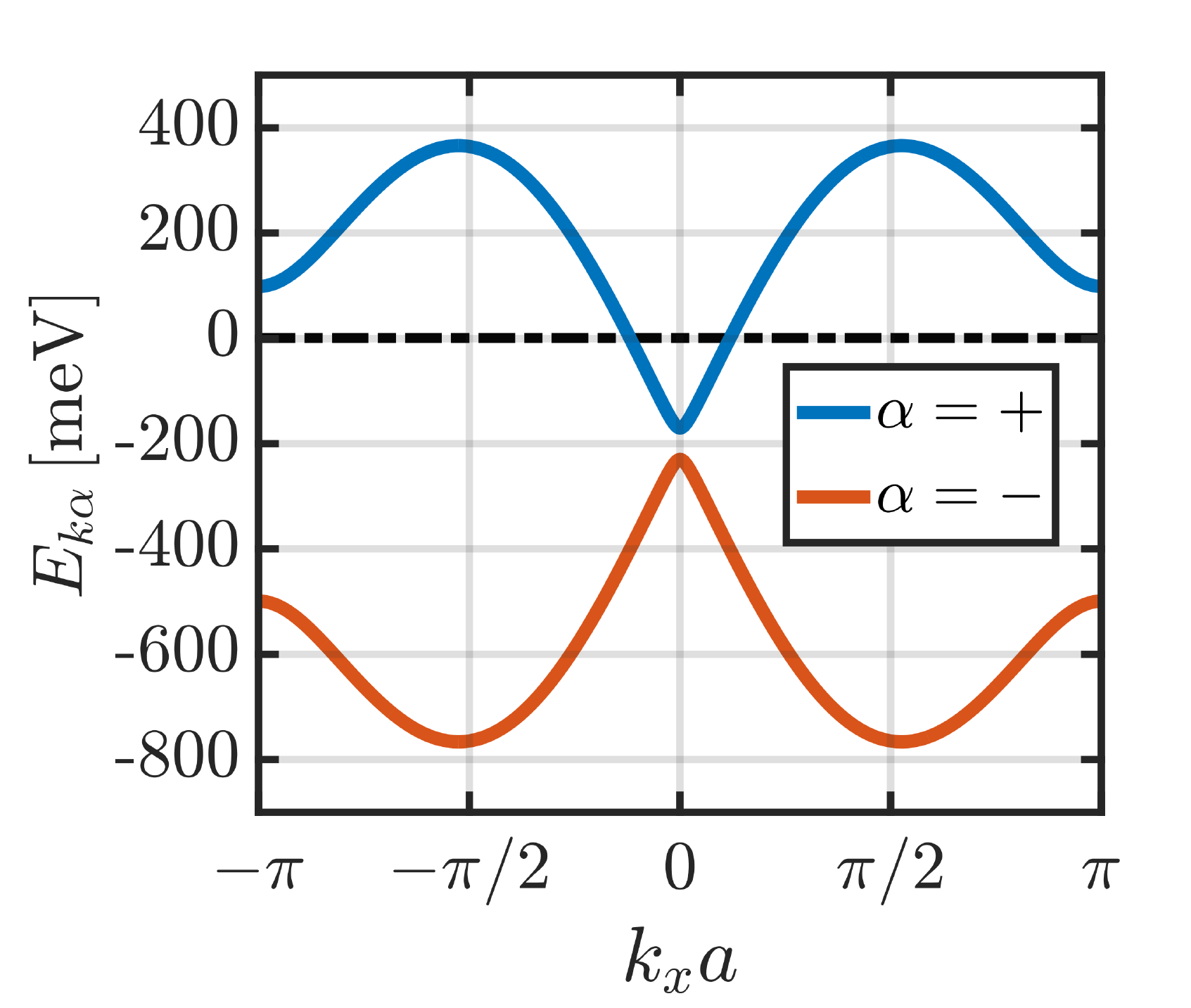}
    \end{center}
    \caption{The band structure of the topological insulator surface states in the presence of the exchange field from the adjacent ferromagnetic insulator for $k_y = 0$, Fermi velocity $v_F = 5\cdot 10^5 \,\rm{m}/\rm{s}$, lattice constant $a = 0.6\,\rm{nm}$, Wilson term coefficient $W = 0.3\,\hbar v_F$ \cite{Zhou2017}, interfacial exchange coupling strength $\Bar{J} = 15 \,\rm{meV}$, spin quantum number of the lattice site spins $s = 1$ and chemical potential $\mu = 200 \,\rm{meV}$. In the long-wavelength limit, the dispersion relation is linear, in agreement with the continuum model. The Wilson terms open a gap at the Brillouin zone boundaries, removing the extra Dirac cones originating with the discretization of the continuum model.}
    \label{fig:TI_band}
\end{figure}

We introduce a Holstein-Primakoff transformation \cite{Holstein1940} for the spin operators $S_{\bm{i}+} = \sqrt{2s}\,a_{\bm{i}}$, $S_{\bm{i}-} = \sqrt{2s}\,a_{\bm{i}}^{\dagger}$, $S_{\bm{i}z}\, = s - a_{\bm{i}}^{\dagger}a_{\bm{i}}$. Including quadratic terms in the magnon operators and performing a Fourier transformation $a_{\bm{i}} = \frac{1}{\sqrt{N}}\sum_{\bm{k}}a_{\bm{k}}e^{-i\bm{k}\cdot\bm{r}_{\bm{i}}}$, the FMI Hamiltonian takes the form \cite{Kittel1963}
\begin{align}
    H_{\rm{FMI}} = \sum_{\bm{k}}\omega_{\bm{k}}a_{\bm{k}}^{\dagger}a_{\bm{k}},
\end{align}
where $\omega_{\bm{k}} = 2sJz_1(1 - \gamma_{\bm{k}}) + 2Ks$ and ${\gamma}_{\bm{k}} = \frac{2}{z_1}\sum_{b}\cos(k_b)$. The number of nearest neighbors has here been denoted by $z_1$, the sum over $b$ covers the spatial dimensions of the FMI lattice, and the number of lattice sites in the interfacial plane is denoted by $N$. \\\indent
From the interaction Hamiltonian \eqref{H_int_F}, we obtain 
\begin{align}
\begin{aligned}
    H_{\rm{int}} = &-2\Bar{J}\sqrt{2s}\,\sum_{\bm{i}}(a_{\bm{i}}\,c_{\bm{i}\downarrow}^{\dagger}c_{\bm{i}\uparrow} + \rm{h.c.})\\ 
    &- 2\Bar{J}\sum_{\bm{i}\sigma}\sigma c_{\bm{i}\sigma}^{\dagger}c_{\bm{i}\sigma}(s - a_{\bm{i}}^{\dagger}a_{\bm{i}}),
\end{aligned}
\end{align}
where the first line originates with the $x,y$-components of the coupling scalar product, and the second line originates with the $z$-omponent. The quantity $\sigma$ in front of the electron operators on the second line is $+1$ for spin up and $-1$ for spin down. Fourier transforming the magnon and electron operators $c_{\bm{i}\sigma} = \frac{1}{\sqrt{N}}\sum_{\bm{k}}c_{\bm{k}\sigma}e^{-i\bm{k}\cdot\bm{r}_{\bm{i}}}$ then produces \cite{Fjaerbu2018}

\begin{align}
\begin{aligned}
           &H_{\text{int}} = V\sum_{\bm{k}\bm{q}}(a_{\bm{q}}c_{\bm{k}+\bm{q}, \downarrow}^{\dagger}c_{\bm{k}\uparrow} + \text{h.c.})\\
           &- 2\Bar{J}s\sum_{\bm{k}\sigma}\sigma c_{\bm{k}\sigma}^{\dagger}c_{\bm{k}\sigma},
\end{aligned}
\end{align}
with $V = \frac{-2\Bar{J}\sqrt{2s}}{\sqrt{N}}$. The first terms represent electron-magnon interactions involving a single magnon, and the second term originates with the exchange field that the electrons on the surface of the TI are exposed to due to the proximity to the FMI. Electron-magnon interactions involving more than one magnon have been neglected. Fourier transforming the electron operators in the TI Hamiltonian, as well as moving the exchange field contribution from the interaction Hamiltonian to the TI Hamiltonian, produces

\begin{align}
\begin{aligned}
        &H_{\text{TI}} = W\sum_{\bm{k}\sigma}\sigma c_{\bm{k}\sigma}^{\dagger}c_{\bm{k}\sigma}\Big[2-\sum_{b}\cos(k_{b})\Big]\\
        &- v_F \sum_{\bm{k}}\Big\{c_{\bm{k}\uparrow}^{\dagger}c_{\bm{k}\downarrow}\Big[\sin(k_y) + i\sin(k_x)\Big] + \text{h.c.}\Big\}\\
        &- 2\Bar{J}s\sum_{\bm{k}\sigma}\sigma c_{\bm{k}\sigma}^{\dagger}c_{\bm{k}\sigma} - \mu\sum_{\bm{k}\sigma}c_{\bm{k}\sigma}^{\dagger}c_{\bm{k}\sigma}.
\end{aligned}
\end{align}
As we are interested in pairing between long-lived excitations on the surface of the TI, mediated by magnetic fluctuations on the surface of the FMI, we diagonalize $H_{\rm{TI}}$, where the presence of the exchange field now has been taken into account. The TI Hamiltonian then takes the form 
\begin{align}
    H_{\text{TI}} = \sum_{\bm{k}\alpha}E_{\bm{k}\alpha}\psi_{\bm{k}\alpha}^{\dagger}\psi_{\bm{k}\alpha},
\end{align}
where $\alpha = \pm$ is the helicity-index of the quasiparticles $\psi_{\bm{k}\alpha}$. Defining $A = -\mu$, $B_{\bm{k}} = W\big[2 - \sum_{b}\cos(k_{b})\big] - 2\Bar{J}s$, $C_{\bm{k}} = -v_F \sin(k_y)$, $D_{\bm{k}} = -v_F \sin(k_x)$, $N_{\bm{k}} = 2F_{\bm{k}}\big(F_{\bm{k}} + B_{\bm{k}}\big)$, and $F_{\bm{k}} = \sqrt{B_{\bm{k}}^2 + C_{\bm{k}}^2 + D_{\bm{k}}^2}$, the excitation energies can be expressed as $E_{\bm{k}\alpha} = -\mu + \alpha F_{\bm{k}}$ and the original electron operators can be related to the new quasiparticle operators
\begin{subequations}
\begin{align}
    c_{\bm{k}\uparrow}   &= Q_{\uparrow +}(\bm{k})\,\psi_{\bm{k}+} + Q_{\uparrow -}(\bm{k})\,\psi_{\bm{k}-},\\ 
    c_{\bm{k}\downarrow} &= Q_{\downarrow +}(\bm{k})\,\psi_{\bm{k}+} + Q_{\downarrow -}(\bm{k})\,\psi_{\bm{k}-},
\end{align}
\end{subequations}
where we have defined
\begin{subequations}
\begin{align}
    Q_{\uparrow +} &=-Q_{\downarrow -}   = (B_{\bm{k}} + F_{\bm{k}})/{\sqrt{N_{\bm{k}}}},\\
    Q_{\uparrow -} &= \,\,\,\,Q^{*}_{\downarrow +}\, = (C_{\bm{k}} + i D_{\bm{k}})/{\sqrt{N_{\bm{k}}}}.
\end{align}
\end{subequations}
The band structure of the TI surface states is presented in Fig.\! \ref{fig:TI_band}, displaying the two bands of opposite helicity. The exchange field from the FMI breaks time-reversal symmetry and introduces a gap in the dispersion relation \cite{Wei2013, Yang2013, Tang2017}, similar to the mass gap in the dispersion relation for massive Dirac fermions \cite{Balatsky2014}. 

Expressing the interaction Hamiltonian in terms of the TI eigen-excitations, the full Hamiltonian becomes $H = H_{\rm{FMI}} + H_{\rm{TI}} + H_{\text{int}}$, with

\begin{align}
\begin{aligned}
    \,\,\,\,\,\,\,\,&H_{\rm{int}} = V\sum_{\bm{k}\bm{q}}\sum_{\alpha\alpha'}Q^{\dagger}_{\downarrow\alpha}(\bm{k+q})Q_{\uparrow\alpha'}(\bm{k})\\
    \,\,\,\,\,\,\,\,&\times a_{\bm{q}}\,\psi_{\bm{k} + \bm{q},\alpha}^{\dagger}\psi_{\bm{k}\alpha'} + \text{h.c.},\\
\end{aligned}
\end{align}
In the following, we derive the effective fermion-fermion interaction arising from this magnon-fermion coupling.

\subsection{Effective interaction}

We proceed by integrating out the magnons in order to obtain an effective theory of interacting helical fermions. Examining the nature of the interaction between particles close to the Fermi surface, we can then determine whether a superconducting instability is possible. We take $H = H_0 + \eta H_1$, where $H_0 = H_{\rm{FMI}} + H_{\rm{TI}}$, $\,\,\eta H_1 = H_{\rm{int}}$ and $\eta$ is a smallness parameter. We then perform a canonical transformation \cite{Kittel1963}

\begin{align}
    H' &= e^{-\eta S}H\,e^{\eta S},
\end{align}

and a second order expansion

\begin{align}
\begin{aligned}
        H' &= H_0 + \eta\Big(H_1 + \comm{H_0}{S}\Big) \\
    &+ \eta^2\Big(\comm{H_1}{S} + \frac{1}{2}\comm{\comm{H_0}{S}}{S} \Big).
\end{aligned}
\end{align}

Choosing 

\begin{align}
\begin{aligned}
    &\eta S \equiv V\sum_{\bm{k}\bm{q}}\sum_{\alpha\alpha'}\Big[x_{\bm{k},\bm{q}}^{\alpha\alpha'}Q^{\dagger}_{\downarrow\alpha}(\bm{k+q})Q_{\uparrow\alpha'}(\bm{k})\,a_{\bm{q}}\\
    &+ y_{\bm{k},\bm{q}}^{\alpha\alpha'}Q^{\dagger}_{\uparrow\alpha}(\bm{k} + \bm{q})Q_{\downarrow\alpha'}(\bm{k})\,a^{\dagger}_{-\bm{q}}\Big]\psi_{\bm{k}+\bm{q},\alpha}^{\dagger}\psi_{\bm{k}\alpha'},
\end{aligned}
\end{align}

and 

\begin{subequations}
\begin{align}
    x_{\bm{k},\bm{q}}^{\alpha\alpha'} &= \frac{1}{E_{\bm{k}\alpha'} - E_{\bm{k} + \bm{q},\alpha} + \omega_{\bm{q}}},\\
    y_{\bm{k},\bm{q}}^{\alpha\alpha'} &= \frac{1}{E_{\bm{k}\alpha'} - E_{\bm{k} + \bm{q},\alpha} - \omega_{\bm{q}}},
\end{align}
\end{subequations}
we have \cite{Schrieffer1966}
\begin{align}
    \eta H_1 + \comm{H_0}{\eta S} = 0,
    \label{condition}
\end{align}
and 
\begin{align}
    H' = H_0 + \frac{1}{2}\comm{\eta H_1}{\eta S}. 
    \label{the_commutator}
\end{align}
Computing the commutator and picking out the terms that involve four fermionic operators, we then obtain

\begin{align}
\begin{aligned}
    &H_{\text{pair}} = V^2\sum_{\bm{k}\bm{q}\bm{k}'}\sum_{\alpha\alpha'}\sum_{\beta\beta'}Q^{\dagger}_{\uparrow\beta}(\bm{k}'-\bm{q})Q_{\downarrow\beta'}(\bm{k}')\\
    &\times Q^{\dagger}_{\downarrow\alpha}(\bm{k+q})Q_{\uparrow\alpha'}(\bm{k})\psi_{\bm{k}+\bm{q},\alpha}^{\dagger}\psi_{\bm{k}'-\bm{q},\beta}^{\dagger}\psi_{\bm{k}'\beta'}\psi_{\bm{k}\alpha'}\\
    &\times \Bigg(\frac{1}{E_{\bm{k}'\beta'} - E_{\bm{k}'-\bm{q},\beta} - \omega_{\bm{q}}} - \frac{1}{E_{\bm{k}\alpha'} - E_{\bm{k} + \bm{q},\alpha} + \omega_{\bm{q}}}\Bigg) .
\end{aligned}
\label{eq:pairing_general}
\end{align}

\begin{figure}[ht] 
    \begin{center}
        \includegraphics[width=0.70\columnwidth,trim= 7.0cm 5.75cm 0.5cm 0.5cm,clip=true]{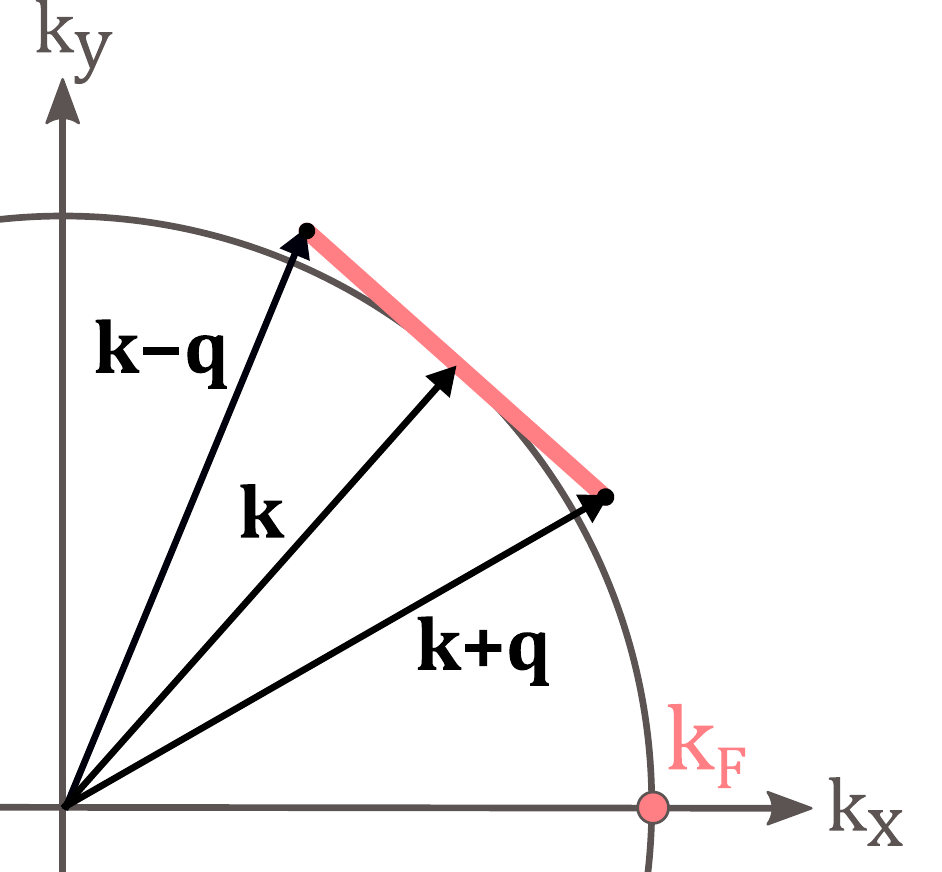}
    \end{center}
    \caption{Amperean pairing with $\bm{k}' \approx \bm{k}$. In order for all momenta to stay close to the Fermi surface, the momentum transfer $q$ needs to be small relative to the Fermi momentum $k_F$.}
    \label{fig:Amperean}
\end{figure}

In the formation of Cooper pairs, the fermions all have momenta close to the Fermi momentum $k_F$. In the case of Amperean pairing \cite{Kargarian2016, Lee2007, Lee2014, Schlawin2019}, where Cooper pairs are formed by particles on the same side of the Fermi surface, this implies that the momentum transfer in the processes, $q$, needs to be small relative to $k_F$ \cite{Hugdal2018}. This is seen in Fig. \ref{fig:Amperean}. In the case of BCS-type pairing between particles on opposite sides of the Fermi surface \cite{Bardeen1957}, $q$ is not necessarily small relative to $k_F$. If $\bm{k} + \bm{q}$ is close to the Fermi surface, then $\bm{k}' - \bm{q}$ is also close to the Fermi surface when $\bm{k}' = -\bm{k}$.

\indent Consider next Eq.\! \eqref{eq:pairing_general} in the long-wavelength limit. Assuming $\mu > 0$, we project down on the helicity band with index $+$, as this is the band that crosses the Fermi level. We then obtain, for BCS-type pairing and Amperean pairing respectively,

\begin{align}
\begin{aligned}
    &H^{(\rm{BCS})}_{\text{pair}} = -\frac{V^2}{4}\sum_{\bm{k}\bm{k}'}\frac{v_F(k_x  - i k_y)}{\sqrt{(2\Bar{J}s)^2 + v_F^2{k}^{2}}}\frac{v_F(k'_x + i k'_y)}{\sqrt{(2\Bar{J}s)^2 + v_F^2{k'}^{2}}}\\
    &\times \frac{2\omega_{\bm{k} - \bm{k}'}}{(E_{\bm{k}',+} - E_{\bm{k},+})^2 - \omega^2_{\bm{k} - \bm{k}'}}\,\psi_{\bm{k},+}^{\dagger}\psi_{-\bm{k}, +}^{\dagger}\psi_{-\bm{k}',+}\psi_{\bm{k}',+},
\end{aligned}
\label{H_FM_BCS}
\end{align}
\begin{align}
\begin{aligned}
    &H^{(\rm{Amp})}_{\text{pair}} = \frac{V^2}{4}\sum_{\bm{k}\bm{k}'\bm{q}}\frac{v_F(k_x - i k_y)}{\sqrt{(2\Bar{J}s)^2 + v^2_F k^2}}\,\frac{v_F(k'_x + i k'_y)}{\sqrt{(2\Bar{J}s)^2 + v^2_F {k'}^2}}\\
    &\times\Bigg(\frac{1}{E_{\bm{k}',+} - E_{\bm{k}' - \bm{q},+} - \omega_{\bm{q}}} - \frac{1}{E_{\bm{k},+} - E_{\bm{k} + \bm{q},+} + \omega_{\bm{q}}}\Bigg)\\
    &\times\psi_{\bm{k}+\bm{q},+}^{\dagger}\psi_{\bm{k}' - \bm{q},+}^{\dagger}\psi_{\bm{k}'+}\psi_{\bm{k}+}.
\end{aligned}
\end{align}
Here we have taken $q \ll k_F$ for the Amperean case. Both of the pairing Hamiltonians include the factors $(k_x - i k_y)(k'_x + i k'_y)$, originating with the spin-momentum locking of the surface states. The denominator of these factors leads to the interaction strength being largest when the Fermi level is far away from the exchange field induced gap in the dispersion relation, $2\bar{J}s \ll v_F k_F$, in agreement with Ref.\! \cite{Hugdal2018}. In the BCS-case, the  factor involving the bosonic and fermionic dispersion relations takes the same form as in phonon-mediated superconductivity \cite{Schrieffer1988}.
\subsection{BCS-type pairing}
For BCS-type pairing, we define 
\begin{align}
\begin{aligned}
    &V^{(\rm{BCS})}_{\bm{k}\bm{k}'} = -\frac{V^2}{2}\frac{v_F(k_x  - i k_y)}{\sqrt{(2\Bar{J}s)^2 + v_F^2{k}^{2}}}\\
    &\times \frac{v_F(k'_x + i k'_y)}{\sqrt{(2\Bar{J}s)^2 + v_F^2{k'}^{2}}} \frac{2\omega_{\bm{k} - \bm{k}'}}{(E_{\bm{k}',+} - E_{\bm{k},+})^2 - \omega^2_{\bm{k} - \bm{k}'}}.
    \label{V_BCS_F}
\end{aligned}
\end{align}
The potential of Eq.\! \eqref{V_BCS_F} can be split into an even and odd part with respect to momentum $V^{(\rm{BCS})}_{\bm{k}\bm{k}'} = V^{(\rm{BCS})}_{\bm{k}\bm{k}',E(\bm{k})} + V^{(\rm{BCS})}_{\bm{k}\bm{k}',O(\bm{k})}$, where $V^{(\rm{BCS})}_{\bm{k}\bm{k}',E(\bm{k})} = (V^{(\rm{BCS})}_{\bm{k}\bm{k}'} + V^{(\rm{BCS})}_{-\bm{k},\bm{k}'})/2$ and $V^{(\rm{BCS})}_{\bm{k}\bm{k}',O(\bm{k})} = (V^{(\rm{BCS})}_{\bm{k}\bm{k}'} - V^{(\rm{BCS})}_{-\bm{k},\bm{k}'})/2$.
By anticommuting the two first fermionic operators and sending $\bm{k} \rightarrow -\bm{k}$, it is seen directly from Eq.\! \eqref{H_FM_BCS} that the even part of the potential vanishes.  Only the odd part can then contribute to pairing. Defining $b_{\bm{k}} = \langle \psi_{-\bm{k},+}\psi_{\bm{k},+}\rangle$ and the superconducting gap $\Delta_{\bm{k}} = - \sum_{\bm{k}'}V^{(\rm{BCS})}_{\bm{k}\bm{k}',O(\bm{k})}b_{\bm{k}'}$, we observe that the superconducting gap function is even in pseudospin (helicity) and odd in momentum, analogous to a spin polarized spin-triplet gap function. The symmetry of the gap function follows directly from the fact that there is only a single band crossing the Fermi level. Performing a standard mean-field procedure \cite{Sigrist1991}, we obtain a self-consistent equation for the superconducting gap function
\begin{align}
    \Delta_{\bm{k}} = - \sum_{\bm{k}'}V^{(\rm{BCS})}_{\bm{k}\bm{k}',O(\bm{k})}\,\frac{\Delta_{\bm{k}'}}{2\Tilde{E}_{\bm{k}'}}\tanh\!\Big(\frac{\beta \Tilde{E}_{\bm{k}'}}{2}\Big),
    \label{FM_BCS_GAP}
\end{align}
where $\Tilde{E}_{\bm{k}} = \sqrt{E^2_{\bm{k}+} + \abs{\Delta_{\bm{k}}}^2}$ and $1/\beta = k_B T$, where $k_B$ is the Boltzmann constant and $T$ is the temperature. While the structure of the interaction potential clearly points to the possibility of a chiral p-wave solution, linearizing the gap equation and performing an average of $V^{(\rm{BCS})}_{\bm{k}\bm{k}',O(\bm{k})}\Delta_{\bm{k}'}$ over the Fermi surface \cite{Sigrist1991}, reveals that the sign of the real part of the potential should have been opposite in order to facilitate this opportunity. An indication of this can be seen directly by inspecting the interaction potential for $\abs{\bm{k}} = \abs{\bm{k}'} = k_F$, producing $\Re\big(V^{(\rm{BCS})}_{\bm{k}\bm{k}',O(\bm{k})}\big) \sim \bm{k}\cdot\bm{k}'$, which is repulsive when $\bm{k}'\parallel\bm{k}$ and attractive when $\bm{k}'\parallel-\bm{k}$. From Eq.\! \eqref{FM_BCS_GAP}, it is clear that the signs of the two sides of the equation will be opposite in these cases. We conclude that BCS pairing is not possible. In accordance with \cite{Kargarian2016}, the interaction potential in the Amperean case appears to have opposite sign of the BCS case, arising from the $Q$-factors generated by the spin-momentum locking of the surface states. We therefore move on to the Amperean case.
\subsection{Amperean pairing}
Taking $\bm{k} = \bm{K} + \bm{p}'$ and $\bm{k}' = \bm{K} - \bm{p}'$, where $\bm{p}' \ll \bm{K}$, and $\bm{p} = \bm{p}' + \bm{q}$ \cite{Lee2007, Kargarian2016, Schlawin2019}, we obtain
\begin{align}
\begin{aligned}
    &H^{(\rm{Amp})}_{\text{pair}} = \frac{1}{2}\sum_{\bm{K}\bm{p}\bm{p}'}V^{(\rm{Amp})}_{\bm{p}\bm{p}'}(\bm{K})\\
    &\times\psi_{\bm{K}+\bm{p},+}^{\dagger}\psi_{\bm{K} - \bm{p},+}^{\dagger}\psi_{\bm{K}-\bm{p}',+}\psi_{\bm{K}+\bm{p}',+},
\end{aligned}
\end{align}
where 
\begin{align}
\begin{aligned}
    &V^{(\rm{Amp})}_{\bm{p}\bm{p}'}(\bm{K}) = \frac{V^2}{2}\frac{v^2_F K^2}{(2\Bar{J}s)^2 + v^2_F K^2}\\
    &\times\Bigg(\frac{1}{E_{\bm{K}-\bm{p}',+} - E_{\bm{K}-\bm{p},+} - \omega_{\bm{p}-\bm{p}'}}\\
    &\hspace{0.2in}- \frac{1}{E_{\bm{K}+\bm{p}',+} - E_{\bm{K} + \bm{p},+} + \omega_{\bm{p}-\bm{p}'}}\Bigg).
\end{aligned}
\end{align}
The momentum $\bm{K}$ should here be located at, or close to, the Fermi surface, restricting the interactions to act between particles located close to the Fermi surface. This potentially leads to the formation of Cooper pairs with center-of-mass momentum $2K \approx 2k_F$. By anti-commuting operators and sending respectively $\bm{p} \rightarrow -\bm{p}$ and $\bm{p}' \rightarrow -\bm{p}'$, we find that the part of the potential that does not vanish is $\Bar{V}_{\bm{p}\bm{p}'}(\bm{K}) \equiv V^{(\rm{Amp})}_{\bm{p}\bm{p}',O(\bm{p},\bm{p}')}(\bm{K}) = (V^{(\rm{Amp})}_{\bm{p}\bm{p}',O(\bm{p})}(\bm{K}) - V^{(\rm{Amp})}_{\bm{p},-\bm{p}',O(\bm{p})}(\bm{K}))/2$, which is odd in both $\bm{p}$ and $\bm{p}'$. Defining $b_{\bm{p}}(\bm{K}) = \langle \psi_{\bm{K} - \bm{p},+}\psi_{\bm{K} + \bm{p},+} \rangle$ and $\Delta_{\bm{p}}(\bm{K}) = - \sum_{\bm{p}'} \Bar{V}_{\bm{p}\bm{p}'}(\bm{K})\,b_{\bm{p}'}(\bm{K})$, we now have a superconducting gap function that is even in pseudospin (helicity) and odd in the relative momentum $\bm{p}$. 

\begin{figure}[ht] 
    \begin{center}
    \hspace{0.0cm}
        \includegraphics[width=0.80\columnwidth,trim= 0.8cm 6.5cm 6.0cm 1.0cm,clip=true]{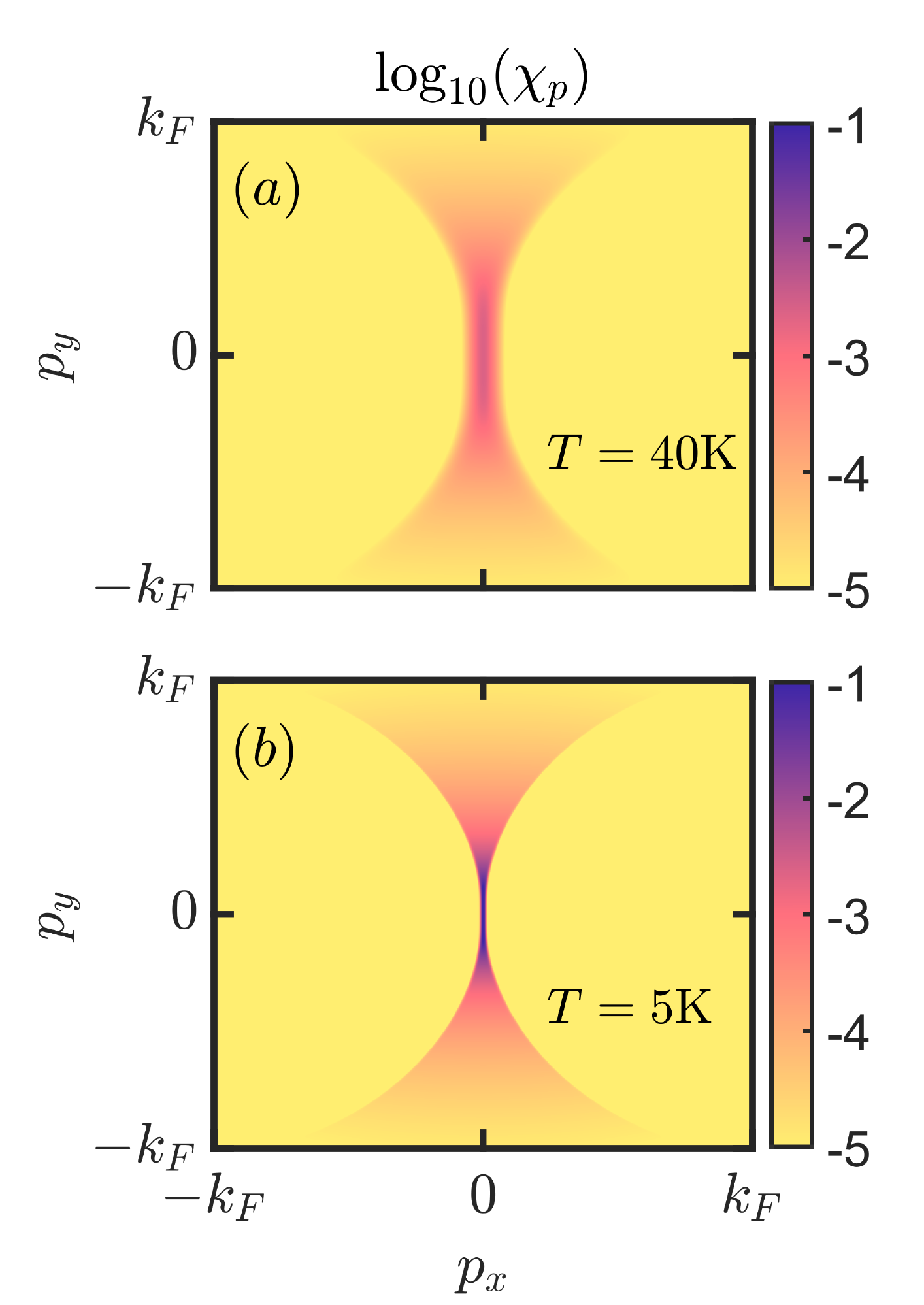}
    \end{center}
    \caption{The factor $\chi_{\bm{p}}(\bm{K})$, plotted on a logarithmic scale as a function of $\bm{p}$ for $\bm{K} = k_F \hat{x}$ for Fermi velocity $v_F = 5\cdot 10^5 \,\rm{m}/\rm{s}$, lattice constant $a = 0.6\,\rm{nm}$, Fermi momentum $k_F a = \pi/8$, interfacial exchange coupling strength $\Bar{J} = 10 \,\rm{meV}$ and spin quantum number of the lattice site spins $s = 1$. In $(a)$, the temperature is set to $T = 40\rm{K}$, while in $(b)$ it is set to $T=5\rm{K}$.}
    \label{fig:chi}
\end{figure} 

Following the mean-field procedure outlined in the Appendix, the self-consistent equation for the gap function takes the form
\begin{align}
    \Delta_{\bm{p}}(\bm{K}) = - \sum_{\bm{p}'}\Bar{V}_{\bm{p}\bm{p}'}(\bm{K}) \Delta_{\bm{p}'}(\bm{K})\chi_{\bm{p}'}(\bm{K}).
    \label{Amp_gap}
\end{align}
Here 
\begin{align}
\begin{aligned}
    &\chi_{\bm{p}'}(\bm{K}) = \frac{1}{4\xi_{\bm{p}'}(\bm{K})}\Bigg[\tanh(\frac{\beta[\xi_{\bm{p}'}(\bm{K}) + \epsilon^{o}_{\bm{p}'}(\bm{K})]}{2})\\
    &\hspace{0.6in}+ \tanh(\frac{\beta[\xi_{\bm{p}'}(\bm{K}) - \epsilon^{o}_{\bm{p}'}(\bm{K})]}{2})\Bigg],
    \label{chi}
\end{aligned}
\end{align}
with the quantities $\epsilon^{o}_{\bm{p}'}(\bm{K}) = (E_{\bm{K}+\bm{p}',+} - E_{\bm{K}-\bm{p}',+})/2$, $\epsilon^{e}_{\bm{p}'}(\bm{K}) = (E_{\bm{K}+\bm{p}',+} + E_{\bm{K}-\bm{p}',+})/2$, and $\xi_{\bm{p}'}(\bm{K}) = \sqrt{[\epsilon^{e}_{\bm{p}'}(\bm{K})]^2 + \abs{\Delta_{\bm{p}'}(\bm{K})}^2}$.

For $\bm{K} = k_F \hat{x}$, which we will focus on in the following, the $\chi$-factor is presented in Fig.\! \ref{fig:chi}, showing that only processes in a small region around $\bm{K}$ give significant contributions to the gap equation. As the Fermi surface is approximately circular, and $\bm{K} = k_F \hat{x}$, it is clear that processes where one of the particles end up on the inside of the Fermi surface are suppressed as temperature is lowered. In order to single out the region of importance for temperatures of the order of a Kelvin, we apply the same ansatz as Kargarian \textit{et al}., $|p_x| < p^2_y/k_F$, originating with Ref.\! \cite{Lee2007}, which is found to be a good approximation.  
\\\indent
For processes where the fermionic quasiparticle energy differences can be neglected, the potential takes the simplified form 
\begin{align}
    \bar{V}_{\bm{p}\bm{p}'}(\bm{K}) \sim \frac{1}{\omega_{\bm{p}+\bm{p}'}} - \frac{1}{\omega_{\bm{p}-\bm{p}'}}.
    \label{nice_pot_FMI}
\end{align}
This potential is attractive for $\bm{p}' \parallel \bm{p}$ and repulsive for $\bm{p}' \parallel -\bm{p}$, which are the signs that are needed in order to obtain a p-wave solution for the gap equation. Note that this is p-wave in the relative momentum. For reasonable parameters, the potential can, however, only be approximated by this form in a very limited region around $\bm{K}$. Outside of this region, the potential changes back and forth between being attractive and repulsive, making it harder to analyze and less favorable for superconductivity. Restricting to the region where $|p_x| < p^2_y/k_F$ and the potential behaves similarly to Eq.\! \eqref{nice_pot_FMI}, assuming that the gap function dies off sufficiently quickly outside of this region, a numerical solution of the linearized gap equation was attempted by picking points in $k$-space within the relevant region and solving the matrix eigenvalue problem using the full potential. The phase space was found to be too small, and the potential not strong enough, to produce a solution, at least not for realistic parameters and reasonable temperatures. The conclusion is therefore that a superconducting instability is not possible within this weak-coupling mean-field theory. According to Kargarian \textit{et al}., a superconducting instability is, on the other hand, possible within a strong-coupling framework. Rather than performing a more advanced analysis of the TI/FMI case, our objective is to compare the results obtained in this section with those of the TI/AFMI case considered in the following section.

\section{Antiferromagnetic case}\label{Section:AFMI_TI}

\begin{figure}[ht] 
    \begin{center}
        \includegraphics[width=0.6\columnwidth,trim= 0.1cm 0.1cm 0.0cm 0.0cm,clip=true]{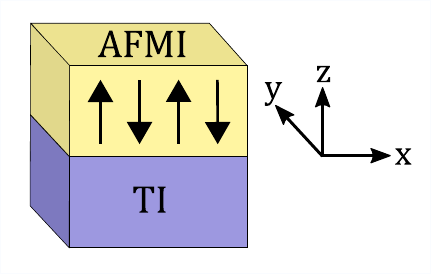}
    \end{center}
    \caption{The system consists of an antiferromagnetic insulator (AFMI) on top of a topological insulator (TI).}
    \label{fig:AFMI_TI}
\end{figure}

In this section, we consider the case of a TI coupled to an AFMI on a bipartite lattice, such as $\rm{Cr_2 O_3}$ or $\rm{Fe_2 O_3}$, as shown in Fig.\! \ref{fig:AFMI_TI}. The interface is once again placed in the xy-plane and the staggered magnetization of the AFMI is assumed to be aligned with the z-direction.

\begin{figure}[ht] 
    \begin{center}
    \hspace{-0.5cm}
        \includegraphics[width=0.8\columnwidth,trim= 1.8cm 0.3cm 3.6cm 0.4cm,clip=true]{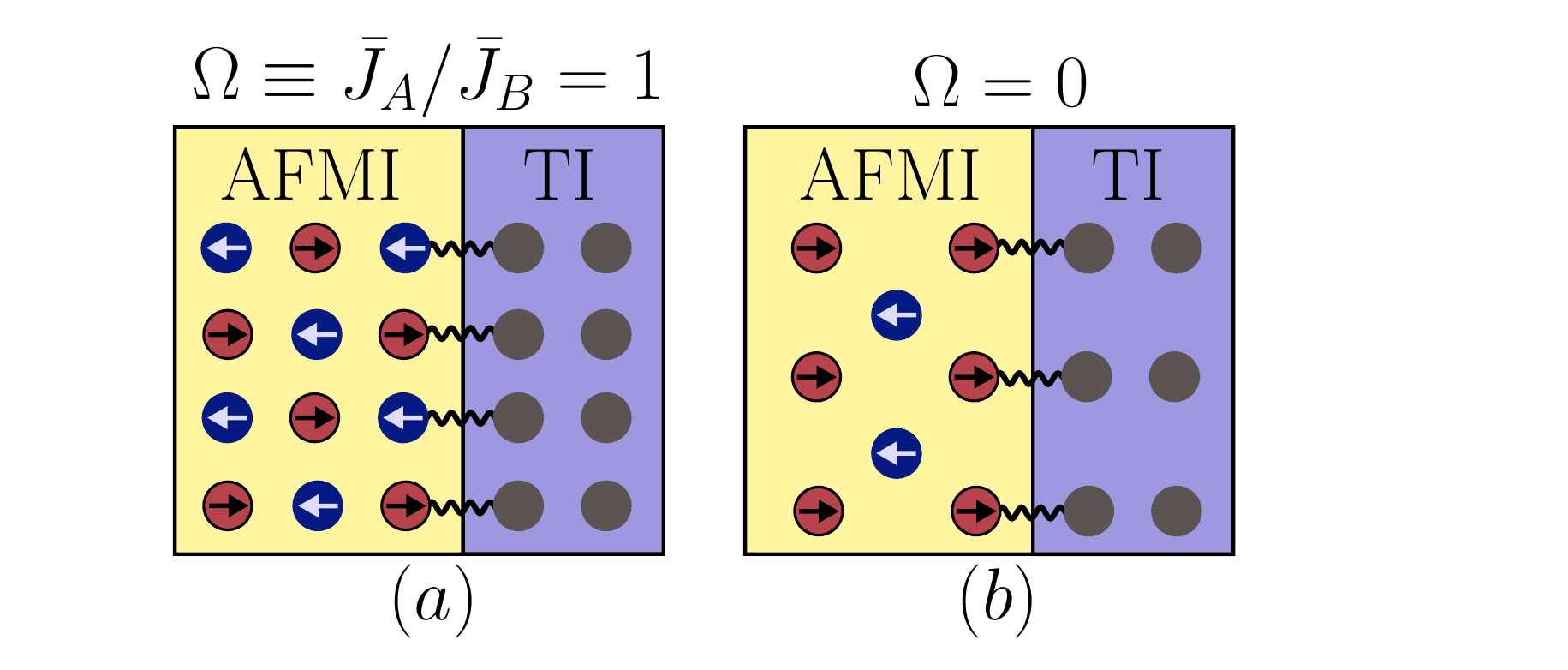}
    \end{center}
    \caption{Exchange coupling across the interface between an antiferromagnetic insulator (AFMI) and a topological insulator (TI). In $(a)$ the antiferromagnetic interface is fully compensated and the TI is coupled symmetrically to the two sublattices of the AFMI with exchange coupling strength $\bar{J}_A = \bar{J}_B$. In $(b)$ the antiferromagnetic interface is fully uncompensated, resulting in the TI coupling to only one of the two sublattices of the AFMI.}
    \label{fig:Coupling}
\end{figure}

\subsection{Model}

The system is modeled by the Hamiltonian \cite{Zhou2017, Li2014}\, $H = H_{\text{AFMI}} + H_{\text{TI}} + H_{\text{int}}$,
\begin{subequations}
\begin{align}
     &H_{\text{AFMI}} = J_1\!\sum_{\langle \bm{i}, \bm{j} \rangle}\! \bm{S}_{\bm{i}} \cdot \bm{S}_{\bm{j}} + J_2\!\sum_{\langle\langle \bm{i}, \bm{j} \rangle\rangle}\!\bm{S}_{\bm{i}} \cdot \bm{S}_{\bm{j}} - K\!\sum_{\bm{i}}S^2_{\bm{i}z},
    \label{H_FM_AF}\\
    &H_{\rm{TI}} = \frac{v_{F}}{2}\sum_{\bm{i},\bm{b}}[c_{\bm{i}}^{\dagger}(i\tau_{y}\delta_{\bm{b}, \Hat{x}} - i\tau_{x}\delta_{\bm{b}, \Hat{y}})c_{\bm{i}+\bm{b}} + \rm{h.c.}] \nonumber\\
   &+ \sum_{\bm{i}}c^{\dagger}_{\bm{i}}[2W\tau_{z} - \mu]c_{\bm{i}} - \frac{W}{2}\sum_{\bm{i},\bm{b}}[c_{\bm{i}}^{\dagger}\tau_z c_{\bm{i}+\bm{b}} + \rm{h.c.}],\label{H_TI_AF}\\
    &H_{\text{int}} =  -2\Bar{J}_A \sum_{\bm{i} \in A} c_{\bm{i}}^{\dagger}\bm{\tau}c_{\bm{i}}\cdot \bm{S_i}  -2\Bar{J}_B \sum_{\bm{i} \in B} c_{\bm{i}}^{\dagger}\bm{\tau}c_{\bm{i}}\cdot \bm{S_i}. 
    \label{H_int_AF}
\end{align}
\end{subequations}
Here, $J_1 > 0$ and $J_2$ are nearest neighbor and next nearest neighbor exchange constants and $K > 0$ parametrizes the easy-axis anisotropy of the AFMI. For $J_2 < 0$, the next nearest neighbor interaction stabilizes the staggered state, while for $J_2 > 0$ this term acts as a frustration. As long as $J_2 > 0$ is small compared to $J_1$, the magnetic ground state is assumed to be an ordered staggered state, while the magnons are influenced by the frustration \cite{Parkinson2010, Richter2010}. The TI Hamiltonian is identical to the one in the previous section and we still consider square lattices. The two subsystems are once again coupled through an exchange interaction, where we now allow for the interaction strength to differ for the A and B sublattices of the AFMI. Such a sublattice-asymmetric interfacial coupling can e.g.\! be achieved using an experimentally realizable uncompensated antiferromagnetic interface \cite{Kamra2017B, Kamra2018}, as depicted in Fig.\! \ref{fig:Coupling}. 

\subsection{Diagonalization of subsystems}
We introduce Holstein-Primakoff transformations for the spin operators on the two sublattices of the AFMI $S_{\bm{i}+}^A = \sqrt{2s}\,a_{\bm{i}}$, $S_{\bm{i}-}^A = \sqrt{2s}\,a_{\bm{i}}^{\dagger}$, $S_{\bm{i}z}^A = s - a_{\bm{i}}^{\dagger}a_{\bm{i}}$, $S_{\bm{j}+}^B = \sqrt{2s}\,b_{\bm{j}}^{\dagger}$, $S_{\bm{j}-}^B = \sqrt{2s}\,b_{\bm{j}}$, and $S_{\bm{j}z}^B = -s + b_{\bm{j}}^{\dagger}b_{\bm{j}}$, and Fourier transformations of the magnon operators $a_{\bm{i}} = \frac{1}{\sqrt{N_A}}\sum_{\bm{k}\in\diamondsuit}a_{\bm{k}}e^{-i\bm{k}\cdot\bm{r}_{\bm{i}}}$, $b_{\bm{i}} = \frac{1}{\sqrt{N_B}}\sum_{\bm{k}\in\diamondsuit}b_{\bm{k}}e^{-i\bm{k}\cdot\bm{r}_{\bm{i}}}$. Here, $\bm{k}\in \diamondsuit$ indicates that the sum covers the reduced Brillouin zone of the sublattices, and the number of lattice sites in the interfacial plane is given by $N = N_A + N_B$. The AFMI Hamiltonian is then diagonalized by a Bogoliubov transformation, expressing the antiferromagnetic eigen-excitations in terms of the original sublattice magnons $\alpha_{\bm{k}} = u_{\bm{k}}a_{\bm{k}} - v_{\bm{k}}b^{\dagger}_{-\bm{k}}$, $\beta_{\bm{k}} = u_{\bm{k}}b_{\bm{k}} - v_{\bm{k}}a^{\dagger}_{-\bm{k}}$. Here, $u_{\bm{k}} = \cosh(\theta_{\bm{k}})$, $v_{\bm{k}} = \sinh(\theta_{\bm{k}})$, $\tanh(2\theta_{\bm{k}}) = -\Tilde{\gamma}_{\bm{k}}/\lambda_{\bm{k}}$, $\Tilde{\gamma}_{\bm{k}} = 4J_1 s \sum_{b}\cos(k_b)$, $\lambda_{\bm{k}} = 2s\big[J_1 z_1 + K + J_2 z_2(\gamma_{\bm{k},2} - 1)\big]$, and $\gamma_{\bm{k},2} = \frac{2}{z_2} \sum_{\substack{\sigma bb' \\ b<b'}}\cos(k_b+\sigma k_{b'})$. The number of next nearest neighbors has here been denoted by $z_2$ and in two dimensions we have $\sum_{\substack{\sigma bb' \\ b<b'}}\cos(k_b+\sigma k_{b'}) = \cos(k_x + k_y) + \cos(k_x - k_y)$. The AFMI Hamiltonian then takes the form 

\begin{align}
    H_{\text{AFMI}} = \sum_{\bm{k}}\omega_{\bm{k}}(\alpha_{\bm{k}}^{\dagger}\alpha_{\bm{k}} + \beta_{\bm{k}}^{\dagger}\beta_{\bm{k}}),
\end{align}
where $\omega_{\bm{k}} = \lambda_{\bm{k}}\sqrt{1-\Tilde{\gamma}^2_{\bm{k}}/\lambda^2_{\bm{k}}}$.\\
\indent For the electron operators, we express the Fourier transformation as $c_{\bm{i}\sigma} = \frac{1}{\sqrt{N}} \sum_{\bm{k}\in \diamondsuit}\big(c_{\bm{k}\sigma}e^{-i\bm{k}\cdot\bm{r}_{\bm{i}}} + c_{\bm{k}+\bm{G},\sigma}e^{-i(\bm{k}+ \bm{G})\cdot\bm{r}_{\bm{i}}}\big)$, where $\bm{G} \equiv \pi(\hat{x} + \hat{y})/a$ is a reciprocal lattice vector for the sublattices. From the interaction Hamiltonian of Eq.\! \eqref{H_int_AF}, we obtain for the two sublattices respectively

\begin{align}
    &H^{(A)}_{\text{int}} = U\,\Omega\sum_{\substack{\bm{k} \in \Box\\ \bm{q} \in \diamondsuit}}\big(a_{\bm{q}}c^{\dagger}_{\bm{k} + \bm{q},\downarrow}c_{\bm{k}\uparrow} + a_{\bm{q}}c^{\dagger}_{\bm{k} + \bm{q} +\bm{G},\downarrow}c_{\bm{k}\uparrow} + \text{h.c.}\big)\nonumber\\
    &- \Bar{J}s\,\Omega\sum_{\substack{\bm{k}\in\Box\\ \sigma}}\sigma\big(c^{\dagger}_{\bm{k}\sigma}c_{\bm{k}\sigma} + c^{\dagger}_{\bm{k}+\bm{G},\sigma}c_{\bm{k}\sigma}\big),\\
    &H^{(B)}_{\text{int}} = U\sum_{\substack{\bm{k} \in \Box\\ \bm{q} \in \diamondsuit}}\big(b_{\bm{q}}c^{\dagger}_{\bm{k} + \bm{q},\uparrow}c_{\bm{k}\downarrow} - b_{\bm{q}}c^{\dagger}_{\bm{k} + \bm{q} +\bm{G},\uparrow}c_{\bm{k}\downarrow} + \text{h.c.}\big)\nonumber\\
    &+ \Bar{J}s\sum_{\substack{\bm{k}\in\Box\\ \sigma}}\sigma\big(c^{\dagger}_{\bm{k}\sigma}c_{\bm{k}\sigma} - c^{\dagger}_{\bm{k}+\bm{G},\sigma}c_{\bm{k}\sigma}\big),
\end{align}
where $\bm{k} \in \Box$ indicates that the sum covers the full Brillouin zone. 
There are additional contributions from two-magnon processes, which we once again neglect. We have here defined $U = -2\Bar{J}\sqrt{s}/\sqrt{N}$, $\Omega \equiv \bar{J}_A / \bar{J}_B$ and $\bar{J} \equiv \bar{J}_B$. The parameter $\Omega$, which is taken as $0 \leq \Omega \leq 1$, then determines the degree of asymmetry in the coupling to the two sublattices of the AFMI. The processes where the momentum of the outgoing electron is shifted by a reciprocal lattice vector $\bm{G}$ are Umklapp processes. These processes are expected to be important for inducing superconductivity mediated by antiferromagnetic magnons in normal metals at half-filling \cite{Fjaerbu2019}. For a tight-binding model, on a square lattice, at half-filling, $\bm{G}$ connects different points on the Fermi surface. For the case of a TI with a Fermi surface close to the center of the Brillouin zone \cite{Xia2009, Chen2009}, on the other hand, the Fermi momentum is typically small compared to $\abs{\bm{G}}$, and these Umklapp processes are expected to be of less importance as they scatter fermions far away from the Fermi surface. The Umklapp processes are therefore neglected in the following.\\
\indent We once again move the exchange field terms, which only cancel for $\Omega = 1$, over to the TI Hamiltonian and express the sublattice magnon operators in the interaction Hamiltonian in terms of the magnons that diagonalized the AFMI Hamiltonian. We then obtain 

\begin{align}
\begin{aligned}
        &H_{\text{int}} = U\sum_{\bm{k}\bm{q}}\Big[\Omega\Big(u_{\bm{q}}\alpha_{\bm{q}} + v_{\bm{q}}\beta^{\dagger}_{-\bm{q}}\Big)c^{\dagger}_{\bm{k}+\bm{q},\downarrow}c_{\bm{k}\uparrow}\\
        &\hspace{0.55in}+ \Big(u_{\bm{q}}\beta_{\bm{q}} + v_{\bm{q}}\alpha^{\dagger}_{-\bm{q}}\Big)c^{\dagger}_{\bm{k}+\bm{q},\uparrow}c_{\bm{k}\downarrow} + \text{h.c.}\Big].
\end{aligned}
\end{align}

\indent For the TI Hamiltonian, we now have
\begin{align}
\begin{aligned}
        &H_{\text{TI}} = W\sum_{\bm{k}\sigma}\sigma c_{\bm{k}\sigma}^{\dagger}c_{\bm{k}\sigma}\Big[2-\sum_{b}\cos(k_{b})\Big]\\
        &- v_F \sum_{\bm{k}}\Big\{c_{\bm{k}\uparrow}^{\dagger}c_{\bm{k}\downarrow}\Big[\sin(k_y) + i\sin(k_x)\Big] + \text{h.c.}\Big\}\\
        &- \Bar{J}s(\Omega - 1)\sum_{\bm{k}\sigma}\sigma c_{\bm{k}\sigma}^{\dagger}c_{\bm{k}\sigma} - \mu\sum_{\bm{k}\sigma}c_{\bm{k}\sigma}^{\dagger}c_{\bm{k}\sigma}.
\end{aligned}
\end{align}
Building on the results from the FMI case, we take $B_{\bm{k}} \equiv W\big[2 - \sum_{b}\cos(k_{b})\big] - \bar{J}s(\Omega - 1)$ and obtain
\begin{align}
    H_{\text{TI}} = \sum_{\bm{k}\alpha}E_{\bm{k}\alpha}\psi_{\bm{k}\alpha}^{\dagger}\psi_{\bm{k}\alpha},
\end{align}
with the rest of the definitions as in the previous section.\\
\indent Expressing the electron operators in the interaction Hamiltonian in terms of the eigen-excitations of the TI Hamiltonian, we obtain

\begin{align}
    &H_{\text{int}}^{(A)} = U\,\Omega\sum_{\bm{k}\bm{q}}\sum_{\alpha\alpha'}\Big[\Big(u_{\bm{q}}\alpha_{\bm{q}} + v_{\bm{q}}\beta^{\dagger}_{-\bm{q}}\Big)\nonumber\\
    &\times Q^{\dagger}_{\downarrow\alpha}(\bm{k+q})Q_{\uparrow\alpha'}(\bm{k})\,\psi_{\bm{k} + \bm{q},\alpha}^{\dagger}\psi_{\bm{k}\alpha'} + \text{h.c.}\Big],\\
    &H_{\text{int}}^{(B)} = U\sum_{\bm{k}\bm{q}}\sum_{\alpha\alpha'}\Big[\Big(u_{\bm{q}}\beta_{\bm{q}} + v_{\bm{q}}\alpha^{\dagger}_{-\bm{q}}\Big)\nonumber\\
    &\times Q^{\dagger}_{\uparrow\alpha}(\bm{k+q})Q_{\downarrow\alpha'}(\bm{k})\,\psi_{\bm{k} + \bm{q},\alpha}^{\dagger}\psi_{\bm{k}\alpha'} + \text{h.c.}\Big].
\end{align}
We will, in the following section, derive the effective fermion-fermion interaction arising from this magnon-fermion coupling.

\subsection{Effective interaction}
We once again perform a canonical transformation in order to obtain a theory of fermions with interactions mediated by magnons. Taking, this time, $\eta H_1 = H^{(A)}_{\text{int}} + H^{(B)}_{\text{int}}$, we choose $\eta S = \eta S^{(A)} + \eta S^{(B)}$ with

\begin{align}
\begin{aligned}
    &\eta S^{(A)} = U\,\Omega\sum_{\bm{k}\bm{q}}\sum_{\alpha\alpha'}\Big[\Big(x^{\alpha\alpha'}_{\bm{k},\bm{q}} u_{\bm{q}}\alpha_{\bm{q}} + y^{\alpha\alpha'}_{\bm{k},\bm{q}}v_{\bm{q}}\beta^{\dagger}_{-\bm{q}}\Big)\\
    &\times Q^{\dagger}_{\downarrow\alpha}(\bm{k+q})Q_{\uparrow\alpha'}(\bm{k}) + \Big(y^{\alpha\alpha'}_{\bm{k},\bm{q}} u_{\bm{q}}\alpha^{\dagger}_{-\bm{q}} + x^{\alpha\alpha'}_{\bm{k},\bm{q}}v_{\bm{q}}\beta_{\bm{q}}\Big)\\
    &\times Q^{\dagger}_{\uparrow\alpha}(\bm{k+q})Q_{\downarrow\alpha'}(\bm{k})\Big]\,\psi_{\bm{k} + \bm{q},\alpha}^{\dagger}\psi_{\bm{k}\alpha'},
\end{aligned}
\end{align}

\begin{align}
\begin{aligned}
    &\eta S^{(B)} = U\sum_{\bm{k}\bm{q}}\sum_{\alpha\alpha'}\Big[\Big(x^{\alpha\alpha'}_{\bm{k},\bm{q}} u_{\bm{q}}\beta_{\bm{q}} + y^{\alpha\alpha'}_{\bm{k},\bm{q}}v_{\bm{q}}\alpha^{\dagger}_{-\bm{q}}\Big)\\
    &\times Q^{\dagger}_{\uparrow\alpha}(\bm{k+q})Q_{\downarrow\alpha'}(\bm{k}) + \Big(y^{\alpha\alpha'}_{\bm{k},\bm{q}} u_{\bm{q}}\beta^{\dagger}_{-\bm{q}} + x^{\alpha\alpha'}_{\bm{k},\bm{q}}v_{\bm{q}}\alpha_{\bm{q}}\Big)\\
    &\times Q^{\dagger}_{\downarrow\alpha}(\bm{k+q})Q_{\uparrow\alpha'}(\bm{k})\Big]\,\psi_{\bm{k} + \bm{q},\alpha}^{\dagger}\psi_{\bm{k}\alpha'},
\end{aligned}
\end{align}
where $x^{\alpha\alpha'}_{\bm{k},\bm{q}}$ and $y^{\alpha\alpha'}_{\bm{k},\bm{q}}$ are defined as in the FMI case.

Computing the commutator in Eq.\! \eqref{the_commutator}, projecting down on the helicity band with index $+$ and taking the long-wavelength limit, we obtain for BCS-type pairing and Amperean pairing respectively

\begin{align}
    \begin{aligned}
    &H^{(\rm{BCS})}_{\rm{pair}} = - \frac{U^2}{4}\sum_{\bm{k}\bm{k}'}\frac{v_F(k_x  - i k_y)}{\sqrt{\big[\bar{J}s(\Omega - 1)\big]^2 + v_F^2{k}^{2}}}\\
    &\times\frac{v_F(k'_x + i k'_y)}{\sqrt{\big[\bar{J}s(\Omega - 1)\big]^2 + v_F^2{k'}^{2}}}\frac{2\omega_{\bm{k} - \bm{k}'}}{\big(E_{\bm{k}',+} - E_{\bm{k},+}\big)^2 - \omega^2_{\bm{k} - \bm{k}'}}\\
    &\times A(\bm{k}-\bm{k}', \Omega)\,\psi_{\bm{k},+}^{\dagger}\psi_{-\bm{k},+}^{\dagger}\psi_{-\bm{k}',+}\psi_{\bm{k}'+},
\end{aligned}
\end{align}

\begin{align}
    A(\bm{q}, \Omega) = \frac{1}{2}(\Omega^2 + 1)(u^2_{\bm{q}} + v^2_{\bm{q}}) + 2\,\Omega\, u_{\bm{q}} v_{\bm{q}},
\end{align}

\begin{align}
    \begin{aligned}
    &H^{(\rm{AMP})}_{\rm{pair}} = \frac{U^2}{4}\sum_{\bm{k}\bm{k}'\bm{q}}\frac{v_F(k_x - i k_y) }{\sqrt{\big[\bar{J}s(\Omega - 1)\big]^2 + v^2_F k^2}}\\
    &\times\frac{v_F(k'_x + i k'_y)}{\sqrt{\big[\bar{J}s(\Omega - 1)\big]^2 + v^2_F {k'}^2}}\Bigg[\frac{1}{2}\Big(\Omega^2 u^2_{\bm{q}} + v^2_{\bm{q}} + 2\,\Omega\,u_{\bm{q}}v_{\bm{q}}\Big)\\
    &\times\Big(\frac{1}{E_{\bm{k}'+}- E_{\bm{k}' - \bm{q},+} - \omega_{\bm{q}}} - \frac{1}{E_{\bm{k}+} - E_{\bm{k} + \bm{q},+} + \omega_{\bm{q}}}\Big)\\
    &+ \frac{1}{2}\Bigg(\Omega^2 v^2_{\bm{q}} + u^2_{\bm{q}} + 2\,\Omega\,u_{\bm{q}}v_{\bm{q}}\Big) \Big(\frac{1}{E_{\bm{k}+} - E_{\bm{k} + \bm{q},+} - \omega_{\bm{q}}}\\
    &- \frac{1}{E_{\bm{k}'+} - E_{\bm{k}' - \bm{q},+} + \omega_{\bm{q}}}\Big)\Bigg]\psi_{\bm{k}+\bm{q},+}^{\dagger}\psi_{\bm{k}' - \bm{q},+}^{\dagger}\psi_{\bm{k}',+}\psi_{\bm{k}+},
\end{aligned}
\end{align}
where we once again have taken $q \ll k_F$ for the Amperean case. 

\begin{figure}[ht] 
    \begin{center}
        \includegraphics[width=0.95\columnwidth,trim= 19.0cm 10.3cm 0.0cm 0.5cm,clip=true]{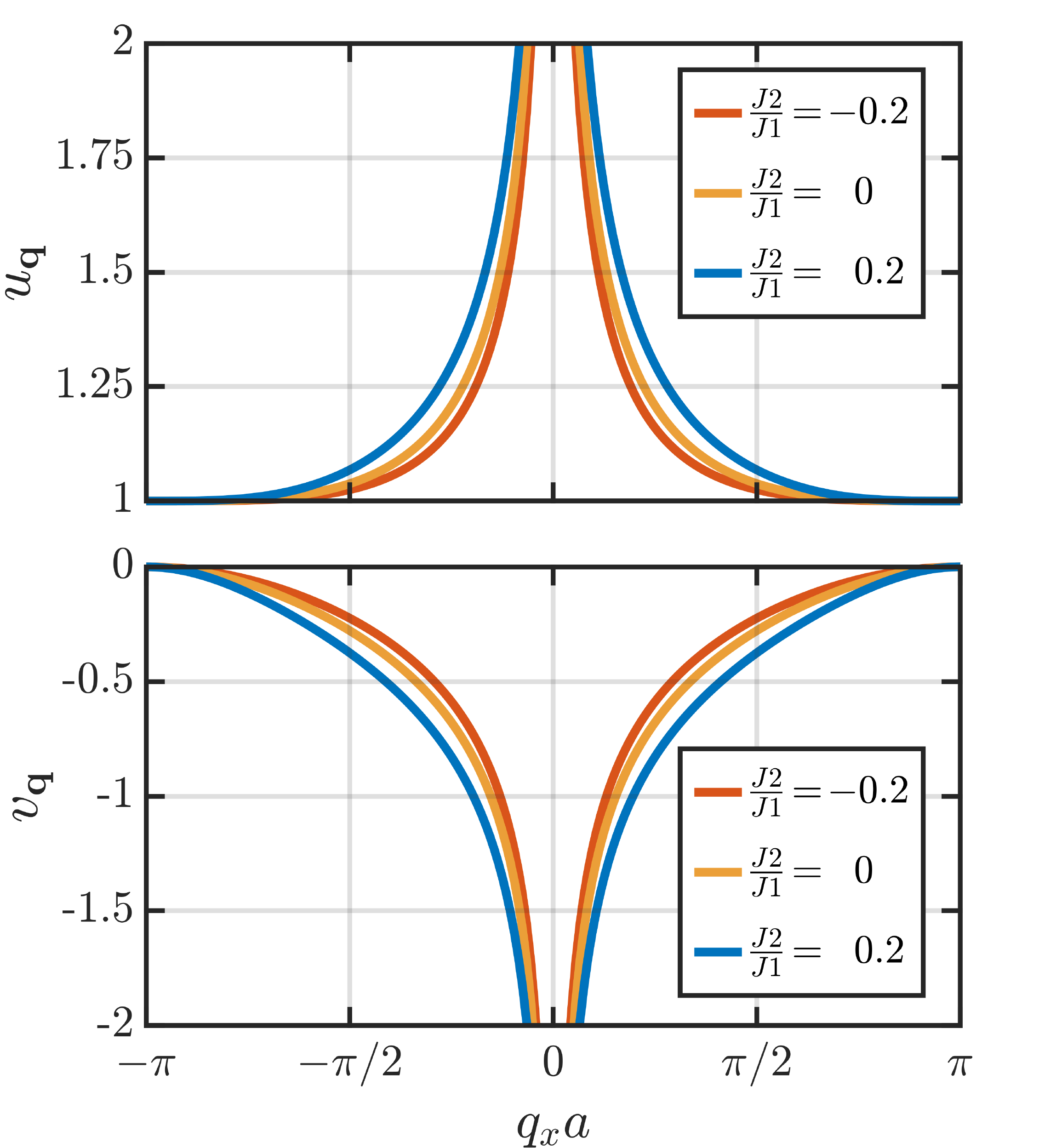}
    \end{center}
    \caption{The magnon coherence factors $u_{\bm{q}}$, $v_{\bm{q}}$ are here presented as a function of the momentum $\bm{q}$ for different values of $J_2/J_1$, where $J_1$ and $J_2$ are the nearest neighbor and next nearest neighbor interaction strengths between the lattice site spins of the antiferromagnetic insulator. We have here set $q_y = 0$, spin quantum number of the lattice site spins $s = 1$, easy-axis anisotropy $K = J_1/10^{4}$ and taken fairly large values for $\abs{J_2/J_1}$ in order to clearly display the effect of the frustration.}
    \label{fig:uv}
\end{figure}

The factor $A(\bm{q}, \Omega)$ is the same as the one arising for asymmetric coupling of a normal metal to the two sublattices of a bipartite AFMI, providing a significant enhancement of the strength of the effective interactions and the superconducting critical temperature in that case \cite{Erlandsen2019}. For long-wavelength magnons, the magnon coherence factors $u_{\bm{q}}$ and $v_{\bm{q}}$ grow large with opposite signs and $A(\bm{q}, \Omega=1) = (u_{\bm{q}} + v_{\bm{q}})^2$ (equal coupling to both AFMI sublattices) is therefore a small quantity while $A({\bm{q}},\Omega=0)=(u_{\bm{q}}^2+v_{\bm{q}}^2)/2$ (only coupling to one AFMI sublattice) is a large quantity \cite{Erlandsen2019}. This enhancement of the interaction for $\Omega = 0$ and suppression for $\Omega = 1$, is a quantum effect not captured by the model in Ref.\! \cite{Hugdal2018}. In the Amperean case, the magnon coherence factors do not combine directly to the same factor $A(\bm{q}, \Omega)$, but, as $u_{\bm{q}}$ and $v_{\bm{q}}$ grow large with opposite signs, the behavior is similar. For $\Omega = 1$, the negative terms $2\,\Omega\, u_{\bm{q}} v_{\bm{q}}$ and positive terms of the form $\Omega^2 u^2_{\bm{q}} + v^2_{\bm{q}}$ work in opposite directions exactly as in $A(\bm{q}, \Omega)$, while for $\Omega = 0$, the negative terms once again vanish, leading to stronger interaction.

\indent The effect of the magnon coherence factors is influenced by the next nearest neighbor interaction term in the AFMI Hamiltonian. In Fig.\! \ref{fig:uv}, the magnon coherence factors $u_{\bm{q}}$ and $v_{\bm{q}}$ are presented as a function of $\bm{q}$ for different values of the next nearest neighbor interaction strength $J_2$. The figure shows that frustrating the system, $J_2 > 0$, increases the magnon coherence factors and thereby also e.g.\! $A({\bm{q}},\Omega=0)=(u_{\bm{q}}^2+v_{\bm{q}}^2)/2$. On the other hand, taking $J_2 < 0$, stabilizing the staggered magnetic state, decreases the magnon coherence factors. This effect would be the same in the normal metal case of Ref.\! \cite{Erlandsen2019}, meaning that next nearest neighbor frustration could aid in the enhancement of the critical temperature also in this case.
\subsection{Pairing}
\indent For the AFMI case, the gap equation for BCS-type pairing is exactly the same as in the FMI case. The only difference between the interaction potentials is the presence of the $A(\bm{q}, \Omega)$ factor in the AFMI case. This factor does not change the sign of the potential and the conclusion is therefore, as for the FMI, that we do not get a chiral p-wave solution to the gap equation.\\\indent 
For Amperean pairing, the AFMI case is also very similar to the FMI case, apart from the presence of the magnon coherence factors. As for the FMI, we obtain a gap equation 
\begin{align}
    \Delta_{\bm{p}}(\bm{K}) = - \sum_{\bm{p}'}\bar{U}_{\bm{p}\bm{p}'}(\bm{K}) \Delta_{\bm{p}'}(\bm{K})\chi_{\bm{p}'}(\bm{K}),
\end{align}
where the potential $\bar{U}_{\bm{p}\bm{p}'}(\bm{K})$ is, once again, odd in both relative momenta. The gap function is, as before, defined as $\Delta_{\bm{p}}(\bm{K}) = - \sum_{\bm{p}'} \Bar{U}_{\bm{p}\bm{p}'}(\bm{K})\,b_{\bm{p}'}(\bm{K})$, and the $\chi$-factor is defined in Eq.\! \eqref{chi}. In a region close to $\bm{K}$, the potential now behaves as 
\begin{align}
    \bar{U}_{\bm{p}\bm{p}'}(\bm{K}) \sim \frac{1}{\omega_{\bm{p}+\bm{p}'}}A(\Omega, \bm{p} + \bm{p}') - \frac{1}{\omega_{\bm{p}-\bm{p}'}}A(\Omega, \bm{p} - \bm{p}').
    \label{nice_pot_AFMI}
\end{align}
We then still have the required signs in order to obtain a non-trivial solution to the gap equation, and obtain a boosting from the $A$-factor, for $\Omega < 1$, in exactly the same way as was obtained in NM/AFMI case of Ref.\! \cite{Erlandsen2019}. For $\Omega = 0$, the interaction potential is therefore much stronger than the potential we had in the FMI case.

\begin{figure}[h!] 
    \begin{center}
    \hspace{0.25cm}
        \includegraphics[width=0.82\columnwidth,trim= 0.1cm 1.3cm 5.5cm 1.05cm,clip=true]{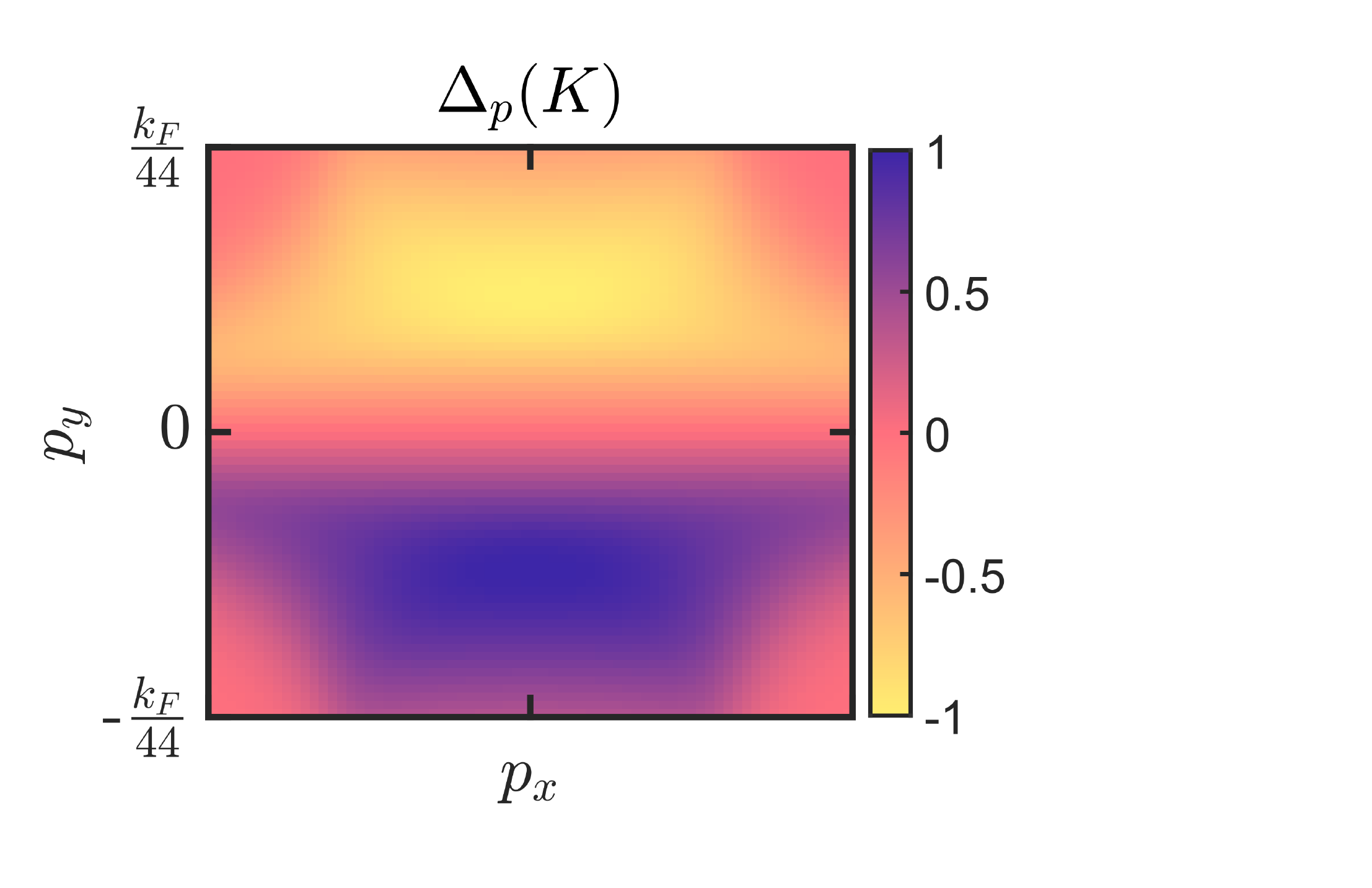}
    \end{center}
    \caption{Form of the gap function obtained as a solution to the linearized gap equation, obtained with Fermi velocity $v_F = 3.6\cdot 10^5 \,\rm{m}/\rm{s}$, lattice constant $a = 0.7\,\rm{nm}$, Fermi momentum $k_F a = \pi/6$, $\bm{K} = k_F \hat{x}$, interfacial exchange coupling strength $\Bar{J} = 18 \,\rm{meV}$, nearest neighbor exchange constant $J_1 = 7 \rm{meV}$, next nearest neighbor exchange constant $J_2 = 0.05 J_1$, easy-axis anisotropy $K = J_1/10^{5}$, spin quantum number of the lattice site spins $s = 1$ and asymmetry parameter $\Omega = 0$. In the x-direction, the points lie within $|p_x| < p^2_y/k_F$, meaning that the $p_x$-value associated with each points depends on the value of $p_y$.}
    \label{fig:gap}
\end{figure} 
\begin{figure}[h!] 
    \begin{center}
    \hspace{-0.58cm}
        \includegraphics[width=0.70\columnwidth,trim= 0.6cm 1.3cm 6.1cm 4.5cm,clip=true]{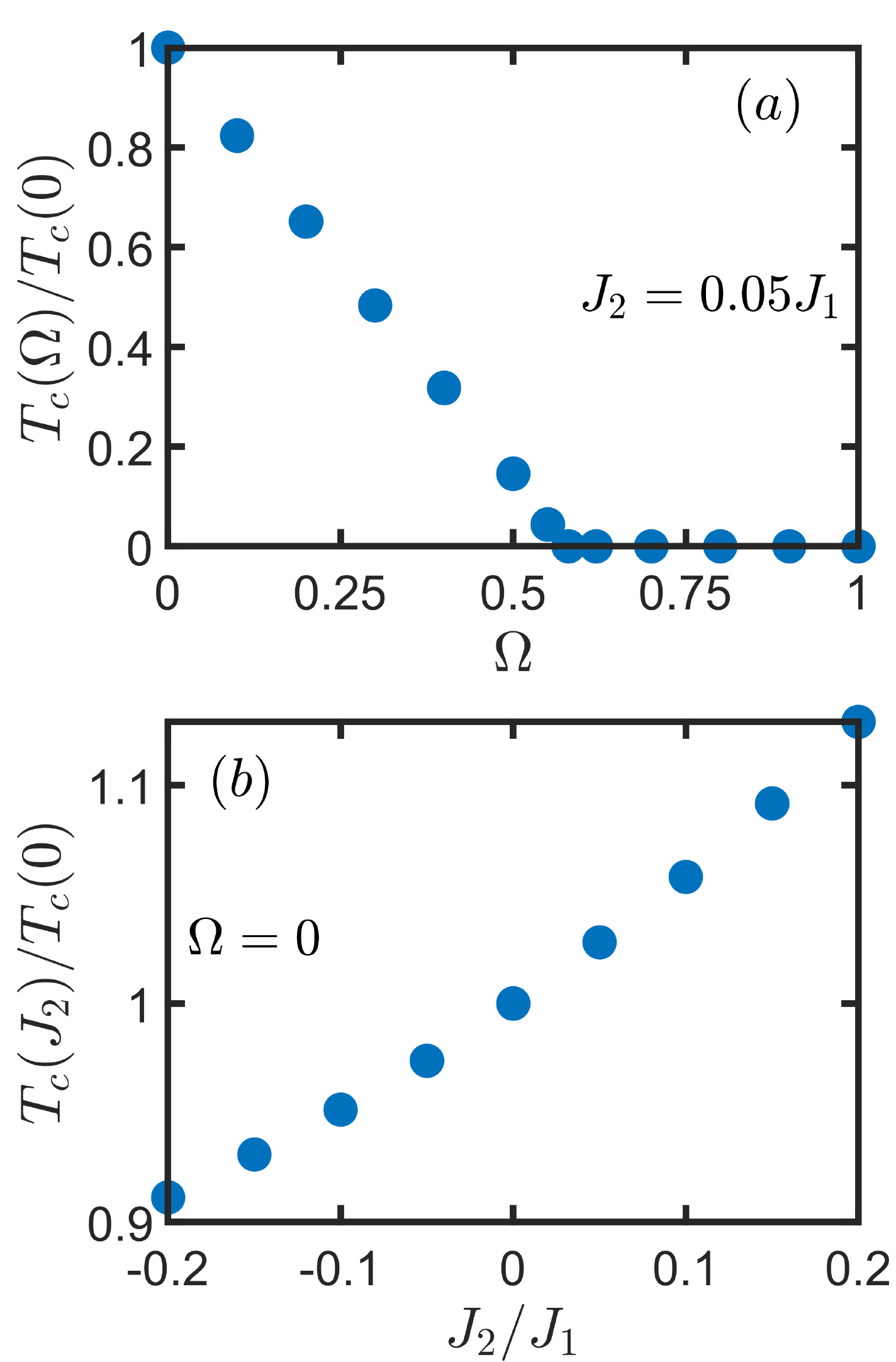}
    \end{center}
    \caption{Critical temperature as a function of $(a)$ asymmetry parameter $\Omega$ and $(b)$ next nearest neighbor exchange constant $J_2$, obtained with Fermi velocity $v_F = 8\cdot 10^4 \,\rm{m}/\rm{s}$, lattice constant $a = 0.7\,\rm{nm}$, Fermi momentum $k_F a = \pi/6$, $\bm{K} = k_F \hat{x}$, interfacial exchange coupling strength $\Bar{J} = 18 \,\rm{meV}$, nearest neighbor exchange constant $J_1 = 7 \rm{meV}$, easy-axis anisotropy $K = J_1/10^{5}$ and spin quantum number of the lattice site spins $s = 1$.}
    \label{fig:Tc}
\end{figure} 

Focusing on $\bm{K} = k_F \hat{x}$, and restricting to the region where $|p_x| < p^2_y/k_F$ and the potential behaves similarly to Eq.\! \eqref{nice_pot_AFMI}, a numerical solution of the linearized gap equation was attempted by picking points in $k$-space within the relevant region and solving the matrix eigenvalue problem using the full potential. The relevant phase space is now typically larger than the corresponding region for the FMI case, as the antiferromagnetic magnons have a linear, instead of quadratic, dispersion relation for small momenta. As the phase space is still small, a strong potential is needed in order to produce a non-trivial solution of the gap equation. Taking $\Omega = 0$ and sufficiently small easy-axis anisotropy, fully exploiting the boosting effect \cite{Kamra2019}, the potential is found to be strong enough to provide a solution. As expected, the solution has a p-wave character, as displayed in Fig.\! \ref{fig:gap}. Since small phase space is compensated by large interaction strength, a strong-coupling approach would provide more solid evidence of the existence of a superconducting instability and realistic estimates for the critical temperature.
\\\indent In order to display the effect of the asymmetry parameter $\Omega$ on the ability of inducing a superconducting instability, we reduce the Fermi velocity to about $20\%$ of typical values \cite{Chen2009} in order to obtain solutions for $\Omega > 0$. The dependence of the critical temperature on $\Omega$ is presented in Fig.\! \ref{fig:Tc} $(a)$, clearly showing that the interaction strength, and thereby the critical temperature is significantly enhanced by coupling the TI asymmetrically to the two sublattices of the AFMI. Similarly, the effect of the next nearest neighbor frustration on the critical temperature is displayed in Fig.\! \ref{fig:Tc} $(b)$. 

\section{Summary}\label{Section:Summary}
We have investigated effective fermion-fermion interactions on the surface of a topological insulator, induced by magnetic fluctuations in a proximity-coupled ferromagnetic or antiferromagnetic insulator. Our main finding is that effective interactions induced by an uncompensated antiferromagnetic interface are significantly stronger than the interactions induced by a fully compensated antiferromagnetic interface or a ferromagnetic interface. This indicates that an uncompensated interface might be the optimal choice for proximity-induced magnon-mediated superconductivity on the surface of a topological insulator. Moreover, we find that the interaction amplification obtained by coupling asymmetrically to the two sublattices of the antiferromagnet can be further strengthened by next nearest neighbor frustration in the antiferromagnet. In both the ferromagnetic and antiferromagnetic cases, we find that the interaction potential has the correct form to give rise to Amperean pairing formed between particles on the same side of the Fermi surface, but in our weak-coupling approach we only find a non-trivial solution of the gap equation in the antiferromagnetic case. 

\section{Acknowledgements}
We thank Niklas Rohling, Akashdeep Kamra, Henning Goa Hugdal, and Even Thingstad for valuable discussions. We acknowledge financial support from the Research Council of Norway Grant No.\! 262633 “Center of Excellence on Quantum Spintronics” and Grant No.\! 250985, “Fundamentals of Low-dissipative Topological Matter”, as well as the European Research Council via Advanced Grant
No.\! 669442, “Insulatronics”.

\appendix
\section{Derivation of Amperean gap equation} \label{App}
We start from the effective Hamiltonian
\begin{align}
\begin{aligned}    
    &H_{\rm{eff}} = \sum_{\bm{k}\alpha} E_{\bm{k}\alpha}\psi^{\dagger}_{\bm{k}\alpha}\psi_{\bm{k}\alpha}\\
    &+\!\sum_{\bm{K}\bm{p}\bm{p}'}\!\frac{\bar{V}_{\bm{p}\bm{p}'}(\bm{K})}{2}\psi_{\bm{K}+\bm{p},+}^{\dagger}\psi_{\bm{K} - \bm{p},+}^{\dagger}\psi_{\bm{K}-\bm{p}',+}\psi_{\bm{K}+\bm{p}',+},
\end{aligned}
\end{align}
where $\bar{V}_{\bm{p}\bm{p}'}(\bm{K})$ is real, odd in $\bm{p}$ and $\bm{p}'$, and $\bar{V}_{\bm{p}\bm{p}'}(\bm{K}) = \bar{V}_{\bm{p}'\bm{p}}(\bm{K})$. Defining $b_{\bm{p}}(\bm{K}) = \langle \psi_{\bm{K} - \bm{p},+}\psi_{\bm{K} + \bm{p},+} \rangle$, we obtain the mean-field Hamiltonian
\begin{align}
\begin{aligned}
    &H'_{\rm{eff}} = \sum_{\bm{k}\alpha} E_{\bm{k}\alpha}\psi^{\dagger}_{\bm{k}\alpha}\psi_{\bm{k}\alpha}\\
    &+\!\sum_{\bm{K}\bm{p}\bm{p}'}\!\frac{\bar{V}_{\bm{p}\bm{p}'}(\bm{K})}{2}\Big[b^{\dagger}_{\bm{p}}(\bm{K})\,\psi_{\bm{K}-\bm{p}',+}\psi_{\bm{K}+\bm{p}',+}\\
    &\hspace{0.87in}+ b_{\bm{p}'}(\bm{K})\,\psi_{\bm{K}+\bm{p},+}^{\dagger}\psi_{\bm{K} - \bm{p},+}^{\dagger}\Big],
\end{aligned}
\end{align}
where terms not affecting for the gap equation have been neglected. Taking 
\begin{align}
\Delta_{\bm{p}}(\bm{K}) = - \sum_{\bm{p}'} \Bar{V}_{\bm{p}\bm{p}'}(\bm{K})\,b_{\bm{p}'}(\bm{K}), 
\label{App_gap}
\end{align}
produces
\begin{align}
\begin{aligned}
    &H'_{\rm{eff}} = \sum_{\bm{k}\alpha} E_{\bm{k}\alpha}\psi^{\dagger}_{\bm{k}\alpha}\psi_{\bm{k}\alpha}\\
    &-\!\frac{1}{2}\sum_{\bm{K}\bm{p}}\Big[\Delta^{\dagger}_{\bm{p}}(\bm{K})\,\psi_{\bm{K}-\bm{p},+}\psi_{\bm{K}+\bm{p},+}\\
    &\hspace{0.4in}+ \Delta_{\bm{p}}(\bm{K})\,\psi_{\bm{K}+\bm{p},+}^{\dagger}\psi_{\bm{K} - \bm{p},+}^{\dagger}\Big].
\end{aligned}
\end{align}
As Cooper pairs with different $\bm{K}$ have different center-of-mass momentum, it is expected that they will, mainly, behave independently of each other. Focusing on a single $\bm{K}$, we can then write 
\begin{align}
\begin{aligned}
    &H'_{\rm{eff}}(\bm{K}) = \sum_{\bm{k}}E_{\bm{k}-}\psi^{\dagger}_{\bm{k}-}\psi_{\bm{k}-} + \frac{1}{2}\sum_{\bm{p}}\begin{pmatrix} \psi^{\dagger}_{\bm{K}+\bm{p},+} & \psi_{\bm{K}-\bm{p},+} \end{pmatrix}\\
    &\times\begin{pmatrix} E_{\bm{K} + \bm{p},+} & -\Delta_{\bm{p}}(\bm{K}) \\ -\Delta^{\dagger}_{\bm{p}}(\bm{K}) & -E_{\bm{K} - \bm{p},+} \end{pmatrix}\begin{pmatrix} \psi_{\bm{K}+\bm{p},+} \\ \psi^{\dagger}_{\bm{K}-\bm{p},+} \end{pmatrix}.
\end{aligned}
\end{align}
The second part of this equation, which is the one relevant for determining the gap equation, can be expressed as
\begin{align}
    H''(\bm{K}) = \frac{1}{2}\sum_{\bm{p}}\phi^{\dagger}_{\bm{p}}(\bm{K})\,M_{\bm{p}}(\bm{K})\,\phi_{\bm{p}}(\bm{K}).
\end{align}
The matrix $M_{\bm{p}}(\bm{K})$ can be transformed into diagonal form by a unitary transformation $P_{\bm{p}}(\bm{K})M_{\bm{p}}(\bm{K})P^{-1}_{\bm{p}}(\bm{K})$, where 
\begin{align}
    P_{\bm{p}}(\bm{K}) = \frac{1}{L_{\bm{p}}(\bm{K})}\begin{pmatrix} \epsilon^{e}_{\bm{p}}(\bm{K}) + \xi_{\bm{p}}(\bm{K}) & -\Delta_{\bm{p}}(\bm{K}) \\ -\Delta^{\dagger}_{\bm{p}}(\bm{K}) & -\epsilon^{e}_{\bm{p}}(\bm{K})-\xi_{\bm{p}}(\bm{K}) \end{pmatrix},
\end{align}
$P^{-1}_{\bm{p}}(\bm{K}) = P_{\bm{p}}(\bm{K})$, $\epsilon^{o}_{\bm{p}}(\bm{K}) = (E_{\bm{K}+\bm{p},+} - E_{\bm{K}-\bm{p},+})/2$, $\epsilon^{e}_{\bm{p}}(\bm{K}) = (E_{\bm{K}+\bm{p},+} + E_{\bm{K}-\bm{p},+})/2$, $\xi_{\bm{p}}(\bm{K}) = \sqrt{[\epsilon^{e}_{\bm{p}}(\bm{K})]^2 + \abs{\Delta_{\bm{p}}(\bm{K})}^2}$, and $L^2_{\bm{p}}(\bm{K}) = 2\xi_{\bm{p}}(\bm{K})\big[\xi_{\bm{p}}(\bm{K}) + \epsilon^{e}_{\bm{p}}(\bm{K})\big]$. We then have
\begin{align}
\begin{aligned}
    H''(\bm{K}) &= \sum_{\bm{p}}\big[\xi_{\bm{p}}(\bm{K}) + \epsilon^{o}_{\bm{p}}(\bm{K})\big]\,\gamma^{\dagger}_{\bm{K} + \bm{p}}\gamma_{\bm{K} + \bm{p}},
\end{aligned}
\end{align}
where the relationship between the original fermionic operators and the $\gamma$-operators is 
\begin{align}
    \psi_{\bm{K} + \bm{p},+} = \frac{\epsilon^{e}_{\bm{p}}(\bm{K}) + \xi_{\bm{p}}(\bm{K})}{L_{\bm{p}}(\bm{K})}\gamma_{\bm{K} + \bm{p}} + \frac{\Delta_{\bm{p}}(\bm{K})}{L_{\bm{p}}(\bm{K})}\gamma^{\dagger}_{\bm{K}-\bm{p}}.
    \label{psi_gamma}
\end{align}
Plugging Eq.\! \eqref{psi_gamma} into the definition of the gap function in Eq.\! \eqref{App_gap}, we obtain Eq.\! \eqref{Amp_gap}. This derivation is similar to the one performed in the supplementary material of Ref.\! \cite{Schlawin2019}.

\bibliographystyle{apsrev4-1}  
\addcontentsline{toc}{chapter}{\bibname}
\bibliography{Refs}  

\end{document}